\documentclass[useAMS,usenatbib]{mn2e}

\usepackage{xcolor}
\usepackage{amsmath}
\usepackage{amssymb}
\usepackage{graphicx,color}
\usepackage{hyperref}

\def\aap{A\&A}
\def\aapr{A\&ARv}
\def\mnras{MNRAS}
\def\apjl{ApJL}
\def\apjs{ApJS}
\def\apj{ApJ}

\def\prd{PRD}

\def\nat{Nature}

\title[Numerical models of BBD-GRBs]{Numerical models of blackbody-dominated gamma-ray bursts -- I. Hydrodynamics and the origin of the thermal emission}
\author[C. Cuesta-Mart\'inez, M. A. Aloy, and P. Mimica]
{C. Cuesta-Mart\'inez$^{1}$\thanks{E-mail: carlos.cuesta@uv.es}, 
M. A. Aloy$^{1}$,
and
P. Mimica$^{1}$\\
$^{1}$Departamento de Astronom\'ia y Astrof\'isica, Universidad de Valencia, 46100 Burjassot (Valencia), Spain}

\pagerange{\pageref{firstpage}--\pageref{lastpage}} \pubyear{}

\begin{document}

\date{\today}

\label{firstpage}

\maketitle

\begin{abstract}
GRB~101225A is a prototype of the class of blackbody-dominated (BBD) gamma-ray bursts (GRBs). It has been suggested that BBD-GRBs result from the merger of a binary system formed by a neutron star and the helium core of an evolved star. We have modelled the propagation of ultrarelativistic jets through the environment left behind the merger by means of relativistic hydrodynamic simulations. In this paper, the output of our numerical models is post-processed to obtain the (thermal) radiative signature of the resulting outflow. We outline the most relevant dynamical details of the jet propagation and connect them to the generation of thermal radiation in GRB events akin to that of GRB~101225A. A comprehensive parameter study of the jet/environment interaction has been performed and synthetic light curves are confronted with the observational data. The thermal emission in our models originates from the interaction between the jet and the hydrogen envelope ejected during the neutron star/He core merger. We find that the lack of a classical afterglow and the accompanying thermal emission in BBD-GRBs can be explained by the interaction of an ultrarelativistic jet with a toroidally shaped ejecta whose axis coincides with the binary rotation axis. The spectral inversion and reddening happening at about 2 d in GRB~101225A can be related to the time at which the massive shell ejected in an early phase of the common envelope evolution of the progenitor system is completely ablated by the ultrarelativistic jet.
\end{abstract}


\section{Introduction}
\label{sec:intro}

Gamma-ray bursts (GRBs) are among the most luminous events in the Universe. They are flashes of gamma-radiation that arrive to Earth from unpredictable directions at random times. GRBs are commonly classified according to their duration: long (LGRB), whose observed duration is longer than 2 s, and short, whose emission lasts less than 2 s \citep{Kouveliotou_etal_1993}. There is a overwhelming observational evidence that LGRBs are formed after the death of massive stars, associated with Type Ic supernovae (SNe) explosions. Nowadays, the paradigm within which we explain the origin of most LGRBs is the collapsar model \citep{Woosley_1993,MacFadyen_Woosley_ApJ_1999__Collapsar}. In this model, a stellar mass black hole (BH) results from the collapse of the massive core of the progenitor star. The BH is surrounded by a thick accretion torus, which is able to produce an ultrarelativistic jet. As has been shown by means of numerical simulations, the jet penetrates the stellar mantle and breaks out through the stellar surface, all the while maintaining a high degree of collimation and low baryon loading (e.g. \citealt{Aloy_etal_ApJL_2000__Collapsar}; \citealt*{Zhang_2003ApJ...586..356Z,Zhang_2004ApJ...608..365Z}; \citealt{Mizuta_etal_2006ApJ...651..960M}; \citealt*{Morsony_etal_2007ApJ...665..569M,Morsony_etal_2010ApJ...723..267M,Mizuta_Aloy_2009,Mizuta_etal_2011ApJ...732...26M}; \citealt{Nagakura_etal_2011ApJ...731...80N}; \citealt*{Nagakura_etal_2012ApJ...754...85N}; \citealt{Lopez_Camara_etal_2013ApJ...767...19L}). 

The interaction of the ultrarelativistic outflow with the circumburst medium causes the formation of external shocks, where highly relativistic electrons produce synchrotron emission, which we observe as an afterglow at frequencies from X-rays down to radio. The spectral energy distribution (SED) of the non-thermal emission produced by either internal or external shocks, as well as their temporal evolution, are commonly characterized by power laws.

In recent years, a handful of GRBs have been discovered, whose properties differ from the standard ones. Among them we point out the superlong and ultralong GRBs (see e.g. \citealt{TS_2005AstL...31..291}; \citealt{Gendre_etal_2013ApJ...766...30G,Levan_etal_2014ApJ...781...13_short}) with durations of about $\sim 10^3$ and $\sim 10^4$\,s. These durations are much longer than those of typical long bursts. Many progenitor scenarios have been proposed for the new discovered bursts, such as tidal disruptions (e.g. \citealt{Lodato_Rossi_2011MNRAS.410..359L}, \citealt{2014arXiv1405.1426M}), core collapse following a long-lasting source (e.g. \citealt{Toma_etal_2007ApJ...659.1420,Nakauchi_etal_2013ApJ...778...67}) or stellar mergers (e.g. \citealt{BK_2010MNRAS.401.1644}; \citealt{Thoene_etal_2011Natur_short}, T11 hereafter).  

For some GRBs associated with SNe, an additional thermal component in the X-ray afterglow has been found and attributed to the SN shock breaking out of the star or the circumstellar wind \citep{Campana_etal_2006Natur_short,Soderberg_etal_2008Natur_short}. Nevertheless, recent observations of bursts, associated with faint SNe, appear with a thermal component not only in X-rays but also in optical bands. This is the case of GRB 101225A (also called the `Christmas burst', CB; T11) which, apart from lasting more than a typical burst, shows an unusually strong blackbody (BB) component in both its X-ray and its optical spectrum. The initial observed duration was longer than 2000\,s. This duration is only a lower bound, since before and after its detection it was out of the field of view of \emph{Swift}. It was also active in the subsequent \emph{Swift} orbit, suggesting a duration in excess of $\sim 7000$\,s (\citealt{Levan_etal_2014ApJ...781...13_short}; L14 hereafter). GRB 101225A does not possess a classical afterglow. Rather, the X-ray emission following the GRB ($0.38$\,h $< t < 18$\,d) is well fitted with an absorbed power-law spectrum in addition to a BB component, i.e. assuming a thermal hotspot with a characteristic temperature $T \sim 1$\,keV (it should be noted that other fits are also possible for describing the early X-ray evolution, see e.g. L14). The ultraviolet--optical--infrared (UVOIR) light curve and SED also display a very peculiar behaviour. During the first 10 d they are best fit as if corresponding to a cooling process of an expanding BB (T11). From the spectral fits to a simple model, which assumes that the observed emission originates from an expanding sphere of a uniform surface temperature, one infers that the radius of the BB-emitting component grows from 13 au ($0.07$ d after the burst) to 45 au ($18$ d after). During the same time interval the BB temperature decreases from $43000$ to $5000$\,K. The radius and temperature evolution of the UVOIR data are radically different from that of the X-ray hotspot, suggesting that they are not caused by the same process. After 10 d there is a flattening of the light curve, suggestive of an associated SN, whose light curve would peak at $\sim 30$\,d. The most recent redshift estimate for the CB is $z=0.847$, and has been obtained by L14 thanks to the identification of [O II], [O III] and H$\beta$ spectral lines. The measured redshift sets a lower bound for the energy of $E_{\rm iso} >~ 1.20 \times 10^{52}$\,erg.

Other bursts, such as GRB 090618 \citep{Page_etal_2011MNRAS.416.2078} or GRB 060218 \citep{Campana_etal_2006Natur_short}, also exhibit a BB component, implying we may be starting to see a new (sub)-class of non-standard GRBs. They are characterized by some thermal element heating the environment, and their central engine may be, in some cases, active for very long time (not the case of GRB 090618). GRB 101225A probably constitutes one of the most prominent examples of the so-called blackbody-dominated GRBs (BBD-GRBs). Thus, it pays off to understand the particularities that differentiate GRB 101225A from other more standard cases. 

In this paper, we perform multidimensional numerical relativistic hydrodynamic (RHD) simulations of jets interacting with an assumed ejecta debris, and further computing the thermal emission from such numerical models. We aim to explain the thermal component observed during the first $5$ d of UVOIR observations that, as we shall see, can be chiefly associated with the jet/ejecta interaction. We will characterize the different thermal signatures to be expected in terms of different physical parameters of our models.  In a companion paper (\citealp{Cuesta_etal_2014_P2}; Paper II hereafter), we specifically focus on trying to understand the complete radiative signature of our models including both, thermal and non-thermal (synchrotron) processes.

The outline of the paper is as follows. In Section~\ref{sec:progenitormodels}, we discuss the possible progenitor scenarios of BBD-GRBs. In Section~\ref{sec:numericalmethod} we set the initial conditions of our hydrodynamic models. In Section \ref{sec:results}, we present the dynamics of the `reference model' (RM; the model that, as it will be seen in Paper II, better explains the observational data) and assess the robustness of our results by considering suitable variations of the main parameters defining the jet, the ejecta and the external medium (EM). The landmark of this paper is presented in Section~\ref{sec:originthermal}, where we show that the  likely origin of the thermal component (observed in the optical observations of GRB 101225A up to the first 5 d) is the interaction between the ultrarelativistic jet and the ejecta. In Section \ref{sec:conclusions}, we summarize the main results of our simulations and discuss the strengths and weaknesses of our models.

\section{Progenitor models}
\label{sec:progenitormodels}

There are two basic alternative progenitor models that could explain (most of) the phenomenology associated with the CB and, by extension, the BBD-GRBs. On the one hand, 
\citet{Nakauchi_etal_2013ApJ...778...67} have proposed that the direct collapse of the envelope of a blue supergiant star may provide the fuel for a very prolonged central engine activity. In this model, the photospheric emission of the cocoon blown by a relativistic jet may be released once the jet breaks out of the surface of the star.  This photospheric emission would create a SN-like bump at longer times, but the spectral inversion of the system at about $\sim 2$ d after the GRB is difficult to explain with such model. A plausible alternative is that the CB resulted from an evolved He star and a neutron star (NS) forming a binary system \citep{FW_1998ApJ,ZhangFryer_2001ApJ...550..357,BK_2010MNRAS.401.1644,BK_2011MNRAS}.

 In this model, a compact object, either a BH or an NS acquires a massive accretion disc by merging with the helium core of its red giant companion. The compact primary enters the helium core after it first experiences a common envelope (CE) phase that carries it inward through the hydrogen envelope (see e.g. \citealt*{Fryer_etal_2006ApJ...647.1269}; \citealt{Chevalier_2012ApJ...752L...2}). The spiral-in process is accompanied by the accretion of several solar masses of helium on a time-scale of minutes and provides a neutrino luminosity, which is very sensitively depending on the mass of the helium core.\footnote{One obtains $L_{\nu\bar{\nu}} \sim 2\times 10^{38} (M_{\rm He}/M_\odot)^{9.5}\,$erg\,s$^{-1}$ for the neutrino--antineutrino annihilation luminosity ($L_{\nu\bar{\nu}}$) of He cores with masses $M_{\rm He}>4 M_\odot$ from the fig.~6 of \cite{Fryer_etal_2013ApJ...764..181}} However, the amount of energy released by neutrinos might not be enough to power a long-lasting jet unless a rather massive He core is invoked, namely, with a mass $\gtrsim 10\,M_\odot$ \citep{Fryer_etal_2013ApJ...764..181}. If the He core mass is relatively small, it is plausible that the BH--disc interaction, mediated by magnetic fields is the primary source of energy of the central engine. In this case, a simplified estimate of the Blandford--Znajek power yields luminosities of $\sim 10^{51}-10^{52}\,$erg\,s$^{-1}$. Longer time-scales to power an outflow shall result if the merger remnant is a magnetar, in which case, the initial NS does not turn into a BH due to an insufficient amount of accreted mass, resulting from a low accretion rate (see, e.g. \citealt{Ivanova_etal_2013_CEEreview_full_list}). NS--He core mergers might occur at a rate comparable to that of merging NSs and BH--NS binaries \citep*{BBF_2012}. The main advantage of this model is that it can account for the observed long duration, and the fact that it provides a simple explanation for the presence of a structured, high-density circumburst environment. The reason is that during the travel of the NS in the CE phase, a fraction of the hydrogen envelope is tidally ejected away from the He core in the form of a thick, dense shell. We refer to it as the CE shell in the rest of this paper. According to recent numerical simulations (e.g. \citealt{Passy_etal_2012ApJ,RT_2012ApJ...746...74,Fryer_etal_2013ApJ...764..181}), the dynamic phase of the CE evolution lasts for 3--5 orbits at the initial binary separation. Taking such a time-scale as a reference, and assuming the debris is ejected at 1--2 times the escape velocity, we can estimate the maximum distance at which it will be located before the merger happens as $R_{\rm debris} \approx (3 - 5) \times t_{\rm orbit} v_{\rm escape} \approx (27 - 45) \times R_{\rm orbit}$, where $R_{\rm orbit}$ is the semimajor axis of the orbit.  The helium-merger model provides a numerically tested explanation for a complex circumburst medium, which roughly resembles a torus or shell located at $\sim 10^{14}$\,cm (which one associates to the debris location after a travel time of $\sim 1.5$\,yr, for an initial orbital separation $R_{\rm orbit} \simeq 30$--$100 R_\odot$, which roughly coincides with the radius of the secondary, evolved, massive star of the binary).  The debris distribution is expected to be non-uniform: most of the mass is ejected along the equator and a low-density funnel is likely to exist, aligned with the rotational axis of the system. Once the two stars merge, an accretion disc and jets are formed leading to a GRB-like event.

In T11, the authors sketched a theoretical model according to which, only a small part of the jet escapes through the funnel giving rise to the detected gamma-ray emission while most of it interacts with the previously ejected material. The outflowing matter interacting with the boundary of the ejecta closer to the rotational axis, leads to a hotspot, producing the persistent X-ray emission. The jet/CE ejecta interaction along the ejecta funnel loads with baryons the relativistic beam of the jet, resulting in a quick deceleration of the jet to mildly relativistic speeds, and thus diminishing any standard afterglow signature. As soon as the jet material breaks out of the shell, it can expand sideways almost freely, forming a hot bubble. The emission from this bubble may account for the UVOIR BB emission before it is finally outshone by the observed SN. 

\cite{Fryer_etal_2013ApJ...764..181} point out that the BB component observed in the CB, in the X-ray flare~060218 and, perhaps, in other low-redshift bursts, is an observational signature of the shell (or torus) of merger ejecta surrounding the burst. The density structure of the medium around the secondary star and, thus, the environment in which the hydrogen envelope is ejected can only be constrained with detailed simulations of He/NS mergers. \cite{ZhangFryer_2001ApJ...550..357} simulations are the only ones which may map the progenitor system that T11 assume for GRB~101225A with sufficiently high numerical resolution. Unfortunately, these models span a typical region of less than $10^{12}\,$cm, and thus, do not include the evolution of the hydrogen envelope. It is not unlikely that, after the SN explosion of the primary (initially most massive) star, the system suffers a `kick' causing that the final merger occurs out of its initial location. Depending on the magnitude of the kick and the time elapsed from the first SN explosion to the final merger, the latter may happen either very close to the original location, or very far away from it. If the merger happens close to its original place, the GRB jet may have to drill its way through the young SN remnant and then through the wind of the secondary star \citep{Fryer_etal_2006ApJ...647.1269}. In the alternative scenario, the merger may take place beyond the termination shock of the wind of the secondary star, in a medium which will be rather uniform if the mass of the secondary star is not too large ($<15$--$20 M_\odot$; \citealt{Fryer_etal_2006ApJ...647.1269}). In any case, the structure of the circumburst medium depends (among other factors) on the exact mass of the He core. As noted in \cite{Fryer_etal_2013ApJ...764..181}, the spectrum of masses of He cores for solar metallicity models spans the range from a few solar masses to a bit more than $25 M_\odot$. If the metallicities are subsolar, He core masses can be found up to $\sim 45 M_\odot$. With such a range of He core masses, the mass of the hydrogen envelope can be rather large (perhaps tens of solar masses), and hence, the amount of mass ejected from the original secondary (massive) star, both prior to the merger with the compact remnant and as a result of it is extremely uncertain. For instance, it is known that sufficiently massive stars (e.g. $\gtrsim 40 M_\odot$) may develop luminous blue variable eruptions during which masses of the order of 1$\,M_\odot$ may be ejected at speeds of the order of the escape velocity of the star. If such a catastrophic event happens a few years before the binary begins its final approach, by the time of merger, this ejection may have travelled a few$\times 10^{15}\,$cm, and filled the environment with mass densities $\sim 10^{-14}\,$g\,cm$^{-3}$. We also point out that late unstable burning may generate gravity waves that deposit their energy and momentum in the outer parts of the star, driving strong mass-loss (e.g. \citealt{Quataert_Shiode_2012MNRAS.423L..92}), and contributing to raise the density of the immediate circumburst medium of the merger. As we shall see, our models accommodate better the observational data if the GRB-jet (true) energy is $\lesssim 10^{52}\,$erg. In order to tap so much energy in the outflow, we need to consider models where the He core mass is larger than $\sim 10 M_\odot$, if the jet is neutrino-powered or $\gtrsim 3 M_\odot$ if it is magnetically powered according to the results of \cite{Fryer_etal_2013ApJ...764..181}.

Another important factor shaping the structure of the medium surrounding the merger is the location of the binary system. It is very likely that a merger among evolved massive stars happens inside of molecular clouds. These molecular clouds may have rather high number densities (e.g. $\sim 10^5\,$cm$^{-3}$ for G353.2+0.9; \citealt{Giannetti_etal_2012A&A...538A..41}, or $\sim 10^7\,$cm$^{-3}$ for the molecular cloud against which Cas A is colliding; \citealt{Fryer_etal_2006ApJ...647.1269}). Therefore, we foresee that the environment surrounding some mergers may be rather massive. As a matter of fact, and as we shall see in Section~\ref{sec:results}, our models suggest that high-density media accommodate better the observational data.
We point out that the interaction of a GRB jet with a dense molecular cloud is not within the scope of our work. We refer the interested reader to \cite{BB_2005ARep...49...24B,BB_2005ARep...49..611B}  for a detailed discussion of the effects that such interaction may have on the resulting optical afterglow of “standard” GRBs. We note that, differently from our model, the optical afterglow in the previous papers results from the reprocessing of the gamma-rays and X-rays produced by the GRB jet in the high-density molecular cloud in which the jet moves (this idea was originally suggested by \citealt{BT_1997AZh....74..483B}).

\section{Numerical method}
\label{sec:numericalmethod}

In order to test the physical model sketched in the previous section, we carry out numerical simulations focusing on the interaction of a relativistic jet with a simple model for the circumburst medium. We have employed the finite volume, high-resolution shock-capturing, RHD code \tiny{MRGENESIS }\normalsize \citep{Aloy_1999ApJS,Mimica_etal_2009ApJ}, in 2D spherical coordinates -- assuming that the system is axisymmetric -- to solve the RHD equations. The code uses a method of lines, which splits the spatial  variation and temporal evolution with two independent discretizations. A Total Variation Diminishing third order Runge--Kutta method and a third order PPM \citep{Colella_Woodward_1984JCoPh..54..174} scheme have been used for both time integration and spatial intercell reconstruction, respectively. Marquina's flux formula \citep{Donat_Marquina_1996JCoPh.125...42} has been chosen for computing the numerical fluxes at the cell interfaces. For the models of interest in this paper, a high order scheme is essential to ameliorate the fine grid needed to resolve both the initial ultrarelativistic jet, as well as the jet/CE-shell interaction. We have produced all our models employing the \emph{TM} approximation \citep*{Mignone_2005ApJS} as an equation of state.

For simplicity, we do not consider general relativistic (GR) effects. This is justified since we begin our jet simulations sufficiently far enough from the central engine of the GRB (where a GR gravitational field is important). In the rest of this paper, we consider flat space--time in the whole numerical domain. Furthermore, magnetic fields are assumed to be dynamically unimportant, so that a pure hydrodynamic approach is used.

In order to compute the thermal emission of our hydrodynamic models and obtain light curves and spectra, we have improved the radiative transport code \tiny{SPEV }\normalsize \citep{Mimica_etal_2004A&A...418..947,Mimica_etal_2005A&A...441..103,Mimica_etal_2009ApJ} to include thermal emission processes (see Appendix \ref{sec:spev}), which can account for the BB component in the observations of GRB 101225A. We assume that the thermal radiation is produced by free--free thermal bremsstrahlung (we also call this model thermal bremsstrahlung-BB). For simplicity in the treatment of the thermal emission, Comptonization is ignored. For the temperatures ($T\la 2\times 10^5$\,K) and number densities ($n\la 10^{14}\,$cm$^{-3}$) of the emitting plasma, thermal bremsstrahlung is the dominant contribution. However, there are (relatively small) emitting regions where Comptonization may be dominant. Ideally, added contributions of both processes should be considered when computing the total thermal emission. However, we are only including one of them, so that our calculation of the thermal emission should be regarded only as a lower bound to the total thermal emission for the proposed model. A rough estimate based on the bolometric power in both free--free bremsstrahlung and Comptonization processes allows us to conclude that the radiative fluxes we compute considering only free--free bremsstrahlung are correct (within a factor $\sim 2$--$3$) during the first 5 d of the system evolution.

To compare with observations we compute the observed flux in the $W2$, $r$ and $X$ bands (corresponding to frequencies of $1.56 \times 10^{15}\,$Hz, $4.68 \times 10^{14}$\,Hz and $2.42 \times 10^{18}$\,Hz, respectively). Because the GRB emission \emph{has been} observed, and due to the probable geometry of the CE shell (with a low-density, narrow funnel along the axis), we assume that the line of sight is aligned with the rotational axis of the system, and that the GRB was observed exactly head-on (i.e. the viewing angle is assumed to be $\theta_{\rm obs} = 0^\circ$).

Since we cannot directly infer from observations all the physical parameters which are necessary to set up the dynamics of an ultrarelativistic jet as well as the environment in which it propagates, we first fix the parameters for the RM, and afterwards we perform a broad parametric scan by varying the properties both of the jet and of the ambient medium. We are not performing consistent numerical simulations of the merger of a He core with a compact remnant. Therefore, we set up the environment and the ejecta debris in an idealized way. For simplicity of the model initialization, we assume that the RM has a uniform medium outside of the He core (which is much smaller than the innermost radius of our computational domain).  We will also alternatively consider external media which are stratified according to different rest-mass density gradients.

\subsection{Setup of the reference model}
\label{sec:setup}

The radial grid of all of our simulations begins at $R_0 = 3 \times 10^{13}$ cm, where an ultrarelativistic jet is injected,\footnote{It is desirable to use a value for $R_0$ as small as possible, to prevent possible numerical `pathologies' related with the start of the jet injection and, later, its switch off. However, much smaller values of $R_0$ than the one we use here would reduce the time step of our models so much to make them computational unfeasible.} and ends at $R_{\rm f} = 3.27 \times 10^{15}$\,cm. It consists of 5400 uniform radial zones.\footnote{Except for the models which are set up with a larger radial grid boundary $R_{\rm f}$ (see Table~\ref{tab:params}), namely M2, G0, S1 and S2, where we use 8500, 8500, 13000 and 18000 radial zones, respectively.} The polar grid spans the range $[0^\circ, 90^\circ]$, with a resolution of 270 uniform zones (i.e. three zones per degree).  We arrived at this particular resolution after performing a convergence study: we performed simulations using progressively finer grids and found that the gross morphodynamical properties of the jet, and the shape of the light curves and spectra have converged (see Appendix \ref{sec:resolution}). Reflecting boundary conditions are imposed at $R_0$, at the rotational axis and at the equator. Outflow boundary conditions are set at $R_{\rm f}$. The grid is initially filled with a cold, static, dense, uniform medium of density $\rho_{\rm ext} = 8 \times 10^{-14}$\,g\,cm$^{-3}$.  With this value of  $\rho_{\rm ext}$, the total mass in the computational domain is $\sim 6 M_\odot$. As we pointed out in Section~\ref{sec:progenitormodels}, this relatively high mass of the environment can only be attained under somewhat extreme conditions, namely, that the progenitor system is embedded in a high-density molecular cloud, and/or that the He star undergoes episodes of violent mass-loss a few years before the merger takes place. In Section~\ref{sec:results}, we will assess more thoroughly how this assumption shapes our main findings by considering lower density environments too.

\subsubsection{CE-shell parameters}
\label{sec:CEshellparams}

\begin{figure}
\centering
\includegraphics[width=8.4cm]{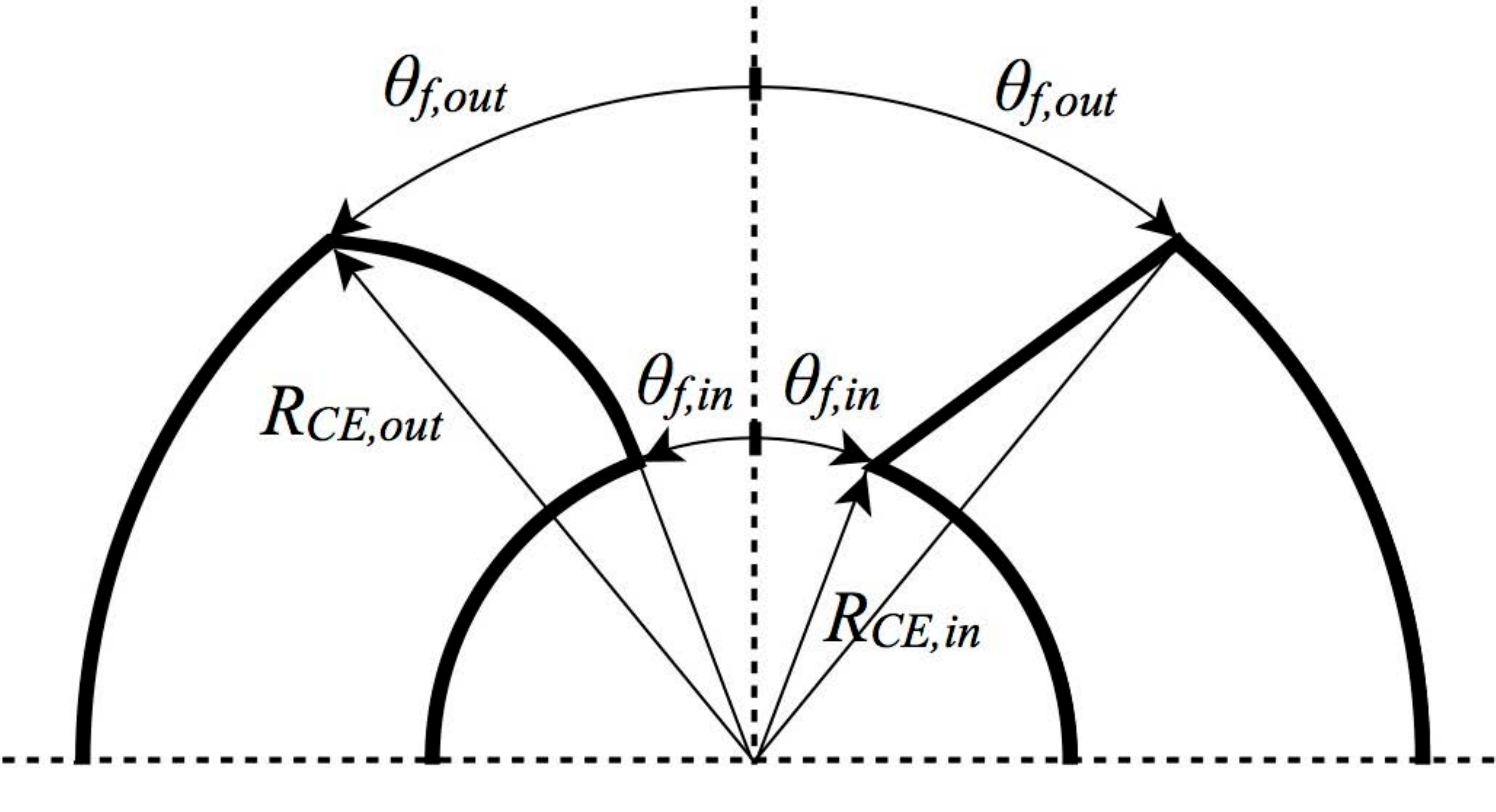}
\caption{Different geometries considered for the CE-shell model: toroidal-like (left) and linear funnel (right). The funnel extends from $\theta_{\rm f,in}$ to $\theta_{\rm f,out}$ (measured, from the rotational axis of the system) in the angular direction, and from $R_{\rm CE,in}$ to $R_{\rm CE,out}$ in the radial direction.}
\label{fig:shellgeometry}
\end{figure}

Starting at a distance $R_{\rm CE,in} = 4.5 \times 10^{13}$\,cm we place a high-density shell that extends out to $R_{\rm CE,out} = 1.05 \times 10^{14}$\,cm (see Fig.~\ref{fig:shellgeometry}). This structure is a simplified model of the ejecta produced during the spiralling of the compact object towards the core of the He star. The gap between the inner radial shell boundary and the radial innermost boundary of the computational domain at $R_0$ is somewhat artificial and its main purpose is to let the jet accelerate smoothly within the computational grid by converting a fraction of its initial thermal energy into kinetic energy (see Section~\ref{sec:jetparameters}). As we will see, this gap has a negligible influence in the resulting light curves, and on the qualitative results we obtain.

 Since the CE shell moves at approximately the escape velocity, its speed is negligible compared to that of any relativistic jet. Thus, we are justified in our assumption that the shell is at rest during the several days that the dynamical jet/shell interaction lasts. The CE shell is uniform in density and pressure, and we assume it is mostly composed of ionized hydrogen ($X_{\rm h}=0.71$). Our model of the CE shell contains a low-density funnel (made of EM) around the symmetry axis with an opening angle of $\theta_{\rm f,in} = 1^\circ$ at $r = R_{\rm CE,in}$. The funnel width grows  exponentially up to $\theta_{\rm f,out} = 30^\circ$ at $r = R_{\rm CE,out}$ to reproduce a toroidal-like shape (see Fig.~\ref{fig:shellgeometry}). In T11 we attributed the X-ray hotspot (observed to be a stationary feature for a few hours after the prompt GRB emission) to the fingerprint of the jet/CE-shell interaction close to the radial innermost boundary of the shell. For such X-ray hotspot, a fixed size of a few $\sim 10^{11}$\,cm would correspond to the transversal radius of the funnel until it is ablated by the jet beam. With the choice of $R_{\rm CE,in}$ and $\theta_{\rm f,in}$ given above, the minimum cross-sectional radius of the funnel is $R_{\rm CE,in} \sin{\theta_{\rm f,in}} \sim 8\times 10^{11}\,$cm. This is somewhat larger than the size of the X-ray hotspot inferred from observational fits. This means that our models will not reproduce the observational signature of the X-ray hotspot very well, since we lack the appropriate numerical resolution (especially in the transversal direction; see Section~\ref{sec:X-rays}).

For the CE-shell density, we take the value $\rho_{\rm CE,sh} = 1.2 \times 10^{-10}$\,g\,cm$^{-3}$, so that  $\rho_{\rm CE,sh} = 1500 \rho_{\rm ext}$ and $p_{\rm CE,sh}/\rho_{\rm CE,sh} \approx 6.7 \times 10^{-9}c^2$ ($c$ being the speed of light in vacuum). This density corresponds to an ejecta mass $\sim 0.26\,M_\odot$. The mass of the CE shell is linked to the mass and metallicity of the secondary star in the binary. However, we do not have exact values for the mass ejection in systems composed of a compact binary and a massive star undergoing to CE phase. From the models of \citet{ZhangFryer_2001ApJ...550..357} it is difficult to estimate the mass ejected from the system beyond a few $10^{11}$\,cm. More massive secondary stars will have larger hydrogen envelopes, which tend to be less gravitationally bound. This means that if the secondary is very massive, one shall expect a large mass in the CE shell. For instance, \citet*{Terman_etal_1995ApJ...445..367T}, who simulate systems with companions of 16 and 24\,$M_\odot$ (with a poor numerical resolution), show that most of the CE shell is ejected during the later spiral-in phase. Nevertheless, it is not unlikely that a sizable fraction of the hydrogen envelope is ejected before the CE-shell phase begins. In this case, the amount of mass left on top of the He core will be the most gravitationally bound part of the envelope and the fraction of such mass tidally ejected could be relatively modest. After an extended numerical experimentation, we consider here relatively low reference values of the CE-shell mass (see Section~\ref{sec:CEshelldensity}).


The pressure in the circumburst medium and in the CE shell is uniform ($p_{\rm ext} = p_{\rm CE,sh}$) and we set it low enough to assure that the plasma is cold (non-relativistic) and has negligible influence on the jet dynamics during the initial 5 d of evolution. We choose   $p_{\rm ext}/\rho_{\rm ext} = 10^{-5}c^2$. At later times the pressure in the cavity blown by the jet decreases until it matches that of the EM. From that point on, the influence of the EM pressure cannot be neglected, but our simulations are stopped well before such pressure matching happens. 

We point out here that, in contrast to \cite*{Badjin_etal_2013MNRAS.432.2454}, who apply a sophisticated radiation transport code to the purpose of estimating the thermal signature of the interaction of both afterglow ejecta and of the prompt radiation emitted by the afterglow ejecta with massive structures in the EM, our shell density ($n_{\rm CE,sh}\sim 10^{14}\,$cm$^{-3}$) is much larger than theirs ($n\sim 10^{10}\,$cm$^{-3}$), and the inner shell radius of our models ($R_{\rm CE,in} = 4.5 \times 10^{13}$\,cm) is much smaller than that of \citeauthor{Badjin_etal_2013MNRAS.432.2454} ($\simeq 10^{16}\,$cm), resulting in rather different physical conditions in the massive shell. In addition, we set ratio $p_{\rm CE,sh}/(\rho_{\rm CE,sh}c^2)$ such that the temperature of the shell is above $10^4$\,K, which allow us to avoid dealing with the possible hydrogen ionization processes in the shell (a microphysical effect which is considered by \citealt{Badjin_etal_2013MNRAS.432.2454}).
\begin{table*}
{\centering
\caption{
Summary of the most important properties that define the different hydrodynamic models in this paper. The equivalent isotropic energy is expressed in units of $10^{53}$\,erg. The rest-mass density contrast $\rho_{\rm CE,sh} / \rho_{\rm ext}$ specified in the third row, refers to the innermost radius ($R_{\rm CE,in}$) of the CE shell. The row `Geometry' refers to the geometrical shape of the CE shell, and models where the shell has a toroidal shape are annotated with `T', and those in which the funnel are linear with `L'. We indicate in bold which parameter of each model is different from RM. We list the innermost radius of the CE shell in units of $10^{13}\,$cm ($R_{{\rm CE,in},13}$). In the penultimate row, models with a uniform EM are annotated with `U' (`U1' denotes a density $\rho_{\rm ext} = 8 \times 10^{-14}$\,g\,cm$^{-3}$ and `U2' a density  $\rho_{\rm ext} = 8 \times 10^{-15}$\,g\,cm$^{-3}$), and models with a stratified medium with `S'. In the last row, we list the outermost radius of our computational domain in units of $10^{15}\,$cm ($R_{{\rm f},15}$).}
\label{tab:params}
\begin{tabular}{|l|c|c|c|c|c|c|c|c|c|c|c|c|c|}
\hline Model & RM & T14 & T20 & E53 & D2  & D3 & GS & G2 & G3 & M2 & G0 & S1 & S2 \\
\hline $\theta_{\rm j}$ & $17^\circ$ & $\boldsymbol{14^\circ}$ & $\boldsymbol{20^\circ}$ & $17^\circ$ & $17^\circ$ & $17^\circ$ & $17^\circ$ & $17^\circ$ & $17^\circ$ & $17^\circ$ & $17^\circ$  & $17^\circ$ & $17^\circ$ \\
          $E_{{\rm iso},53}$ & 4 & 4 & 4 & \textbf{2} & 4 & 4 & 4 & 4 & 4 & 4 & 4 & 4 & 4\\
          $\rho_{\rm CE,sh} / \rho_{\rm ext}$ & 1500 & 1500 & 1500 & 1500 & \textbf{817}  & \textbf{15000} & \textbf{4304} &1500 &1500& \textbf{15000}& \textbf{15000}  &  1500 & 1500 \\ 
          Geometry & T & T & T & T & T & T & \textbf{T}$^{\rm a}$ & \textbf{L} & T &  T & T & T & T \\	
          $\theta_{\rm f,out}$ & $30^\circ$ & $30^\circ$& $30^\circ$& $30^\circ$& $30^\circ$& $30^\circ$& $30^\circ$& $30^\circ$ & $\boldsymbol{15^\circ}$& $30^\circ$ & $30^\circ$ &$30^\circ$ & $30^\circ$ \\
          $R_{\rm CE,in,13}$  & 4.5 & 4.5 & 4.5 & 4.5 & 4.5 & 4.5 & 4.5 & 4.5 & 4.5 & 4.5 &{\bf3} & 4.5 & 4.5\\		
          Ext. medium & U1 & U1 & U1 & U1 & U1 & U1 & U1 & U1 & U1 &  \textbf{U2}$^{\rm b}$ & \textbf{U2}$^{\rm b}$ & \textbf{S}$^{\rm c}$& \textbf{S}$^{\rm d}$ \\
          $R_{{\rm f},15}$ & 3.27 & 3.27 & 3.27 & 3.27 & 3.27 & 3.27 &3.27 &3.27 &3.27 & {\bf 5.13} & {\bf 5.13} & {\bf 7.83} & {\bf 10.83} \\
\hline
\end{tabular}
}
{\\$^{\rm a}$ In model GS, the CE-shell rest-mass density and pressure are not uniform but decay with $r^{-2}$.
\\$^{\rm b}$ In the models M2 and G0, the pressure in the EM and CE shell is $p_{\rm ext}/\rho_{\rm ext} = 10^{-4}c^2$.
\\$^{\rm c}$ In model S1, the EM has a rest-mass density and pressure that decay with $r^{-1}$.
\\$^{\rm d}$ In model S2, the EM has a rest-mass density and pressure that decay with $r^{-2}$.
}
\end{table*}

\subsubsection{Jet parameters}
\label{sec:jetparameters}
\begin{figure*} 
\centering \includegraphics[width=13.5cm]{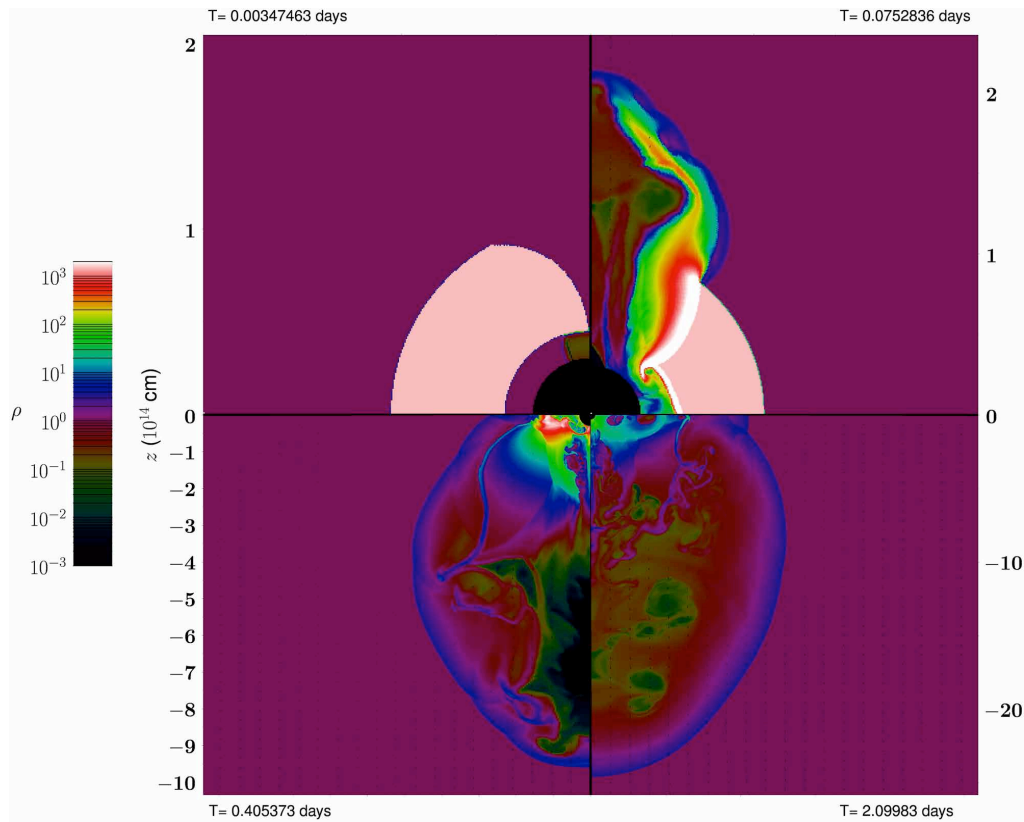} 
\caption{Four snapshots of the rest-mass density evolution of the RM. The rest-mass density is normalized to the EM density $\rho_{\rm ext} = 8 \times 10^{-14}$\,g\,cm$^{-3}$. The time displayed in each panel refers to the laboratory frame time. In the upper-left panel, we show the geometry of the shell before the jet impacts it. The upper-right panel shows the jet penetrating the shell. The lower-left panel displays the jet having developed a quasi-spherical bubble after interacting with the shell. Finally, in the lower-right panel it can be seen the self-similar expansion of the hot bubble through the circumburst medium after $\sim 2$\,d.}
\label{fig:hydro-RM-evo}
\end{figure*}

For the RM, we have chosen a jet opening angle of $\theta_{\rm j} = 17^\circ$ (i.e. $\theta_{\rm j} \gg \theta_{\rm f,in}$), ensuring that the beam of the jet spans a wedge wider than the funnel when it hits the innermost radial boundary of the CE shell. The jet has an initial Lorentz factor $\Gamma_i = 80$, and its specific enthalpy is set to $h_{\rm i} = 5$, so that it can potentially accelerate to an asymptotic Lorentz factor $\Gamma_\infty \approx 400$, by virtue of the relativistic Bernoulli's law. Indeed, we set up the inner `gap' between the CE shell and the jet injection nozzle (see previous section) for numerical convenience. In this way, we let the jet to speed up smoothly and within the grid to Lorentz factors above 100 before it collides with the CE shell. In order to set a reference value for the total jet energy we consider that it is constrained by the observed lower bound of $E_{\rm iso,\gamma+X} > 1.2 \times 10^{52}$\,erg (L14). The isotropic equivalent total energy of the jet will be larger than this value, since we do not exactly know the radiative efficiency in gamma- and X-rays of the jet ($\epsilon_{\rm R}$), namely, $E_{\rm iso,\gamma+X}=\epsilon_{\rm R} E_{\rm iso}$. We choose $E_{\rm iso} = 4 \times 10^{53}$\,erg for our reference model. This means that our reference jet model has a true energy $E_{\rm j}= E_{\rm iso}(1-\cos{\theta_{\rm j}})/2 = 8.7\times 10^{51}$\,erg, which is a likely fraction of the available rotational energy (few times $10^{52}\,$erg) if the central engine is a protomagnetar \citep{Metzger_2011MNRAS.413.2031}. We also note that this jet energy could be on reach in neutrino-powered jets if the mass of the He core is sufficiently large \citep{Fryer_etal_2013ApJ...764..181}.

On the other hand, observations provide us with a lower bound for the burst duration. We take it as a reference for setting the total injection time, $t_{\rm inj} = 7000$\,s (see L14).  Also, as we can see from the fig. 1 of the supplementary material in T11, the jet injection luminosity can be assumed to be constant only up to $t_1 = 2000$\,s, and then decreasing until $ t_2 = t_{\rm inj}$. More specifically, and taking into account that when setting-up a jet a suitable transformation of $t_1$ and $t_2$ to the laboratory frame (attached to source) shall be done (namely, $T = t / (1 + z)$), we consider a two-phase injection:\footnote{In \cite{Aloy_etal_2013ASPC..474...33} we assumed a unique constant injection interval with duration equal to $t_{\rm inj}$.} (1) constant up to $T_1$ and (2) variable (with a dependence $t^{-5/3}$, similar to that expected from tidal disruption events up to $T_2$. With the known redshift $z=0.847$ we obtain $T_1 \simeq 1100$\,s and $T_2 \simeq 3800$\,s. It is numerically convenient for $T>T_2$ to progressively switch off the jet by reducing both the injected rest-mass density and pressure as $\propto t^{-4}$, rather than switching it abruptly off.
\begin{figure*}
\centering
\includegraphics[width=17.6cm]{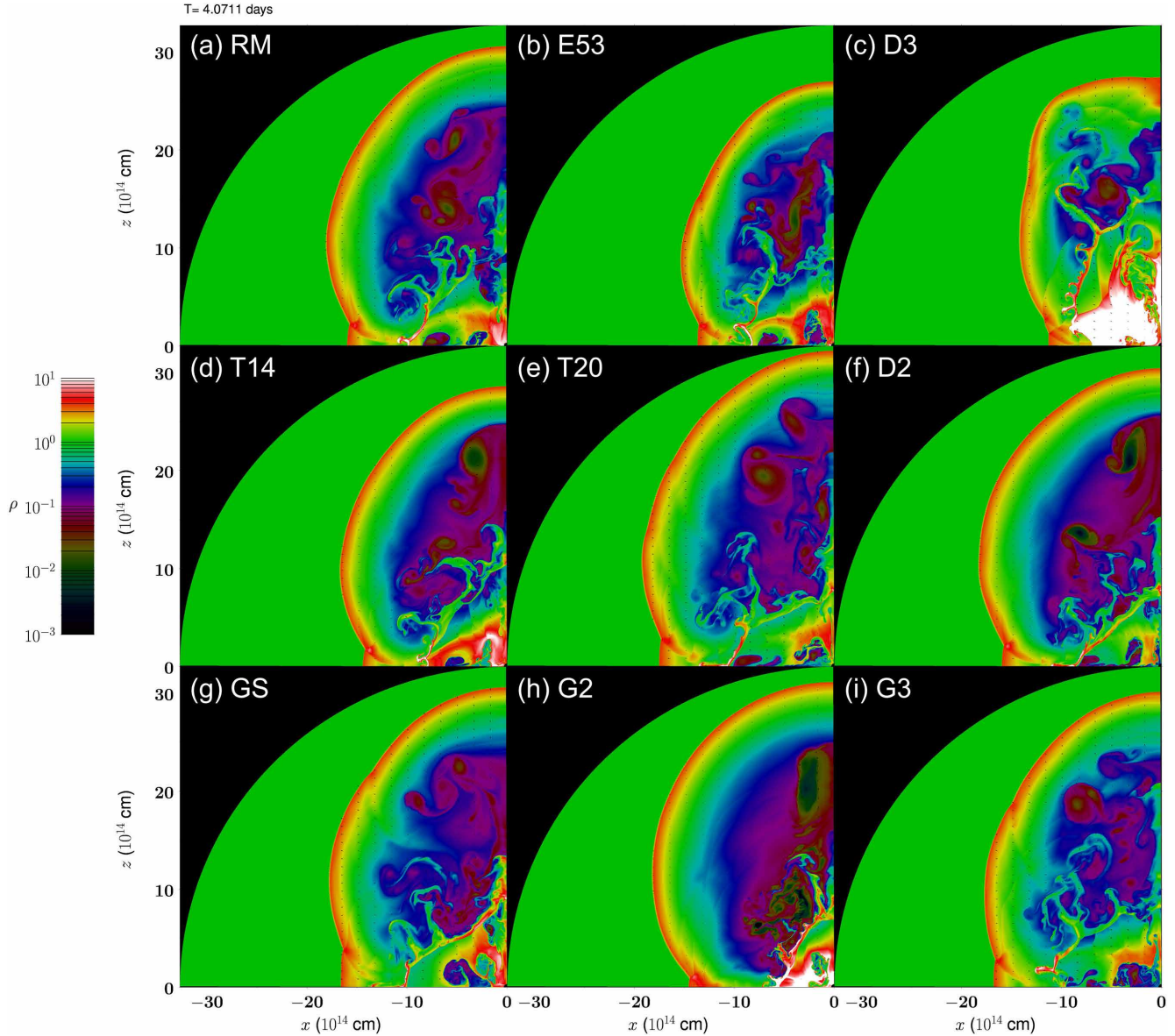}
\caption{Snapshots of the rest-mass density of all the models with a uniform, high-density EM at the end of the computed evolution  ($T= 4.0711$\,d): (a) RM, (b) E53, (c) D3, (d) T14, (e) T20, (f) D2, (g) GS, (h) G2 and (i) G3 (see Table~\ref{tab:params}).}
\label{fig:hydro-models}
\end{figure*}
%

\section{Results}
\label{sec:results}

In this section, we describe the morphology and the dynamics of the RHD jet simulations performed with \tiny{MRGENESIS}\normalsize. We first discuss the results for the RM (Section~\ref{sec:referencemodel}) and then consider variations of the parameters in Section~\ref{sec:parametricscan}. A summary of the most salient parameters of the models presented here is given in Table~\ref{tab:params}, where also the names of each of the models are listed.

\subsection{Reference model}
\label{sec:referencemodel}

In Fig.~\ref{fig:hydro-RM-evo}, we show four snapshots of the RM evolution. Shortly after the start of the jet injection, within the first few seconds, the jet starts to hit the inner boundary of the CE shell (Fig.~\ref{fig:hydro-RM-evo}, upper-left panel).  As a result a pair of shocks form that rapidly heat the plasma to temperatures of up to $\sim {\rm  few} \times 10^6$\,K. The properties of these shocks are not the standard ones expected for the forward and reverse shocks in relativistic ejecta associated with GRB afterglows. Instead, they are propagating at Newtonian speeds, starting at the funnel walls and moving laterally towards the jet axis. In the process, the shocks are also penetrating the CE shell and moving sideways, in a direction almost perpendicular to the jet propagation and, hence, to the line of sight (the shock can be seen as white shades in Fig.~\ref{fig:hydro-RM-evo}, upper-right panel).

During the time in which we keep the jet injection conditions through the inner boundary of our computational domain, a fraction of the jet close to the axis (its innermost core) flows with a negligible resistance. However, the jet is broader than the narrow CE-shell funnel and, hence, a major fraction of the jet volume impacts on the inner radial edge of the CE shell. Since the CE shell is much denser than the jet, the result of the CE-shell/jet interaction is the jet baryon loading, which quickly (within hours) decelerates it to subrelativistic speeds. After about 0.1\,d, most of the mass of the CE shell originally located in the angular region $[\theta_{\rm f,in},\theta_{\rm j}]$ is incorporated into the jet beam and surrounding cocoon. 

The subsequent jet evolution is determined by the balance between the injected jet energy and the mass ploughed by the cavity blown by the jet from the EM. As we shall see, all models propagating into a uniform circumstellar medium pile up $\sim 1$--$2\,M_\odot$ of EM and tend to develop a spherical shape in the long term. 

\subsection{Parametric scan}
\label{sec:parametricscan}

We have presented the RM as a prototype of the evolution of an ultrarelativistic jet piercing a massive shell that results from the ejection of the envelope of the stellar progenitor. In the following, we will show how changes in the assumed parameters of our models shape the resulting dynamics  and also we will assess the robustness of the results. Along the way,  we will show that the generic long-term evolution of all the models we have explored is such that they behave almost self-similarly.  In Table \ref{tab:params}, we show the parameters of the RM subject to variation in the parametric scan. The rest of the models are produced changing only one or two of the parameters with respect to the RM. On the basis of these results, we will assess the origin of the thermal emission in Section~\ref{sec:originthermal}.

\subsubsection{Isotropic energy of the jet, $E_{\rm iso}$}

We have evolved two models with different isotropic energies $E_{\rm iso} = 4 \times 10^{53}$ (RM, Fig.~\ref{fig:hydro-models}a) and $2\times10^{53}$ erg (E53, Fig.~\ref{fig:hydro-models}b). The rest of the parameters are the same, especially the jet half-opening angle, $\theta_{\rm j} = 17^\circ$. The size of the bubble blown by the jet is proportional to the equivalent isotropic energy of the models. We see that bubbles in more energetic models propagate faster, reach larger distances, and have a more spherical shape. 
\begin{figure}
\centering
\includegraphics[height=5.7cm]{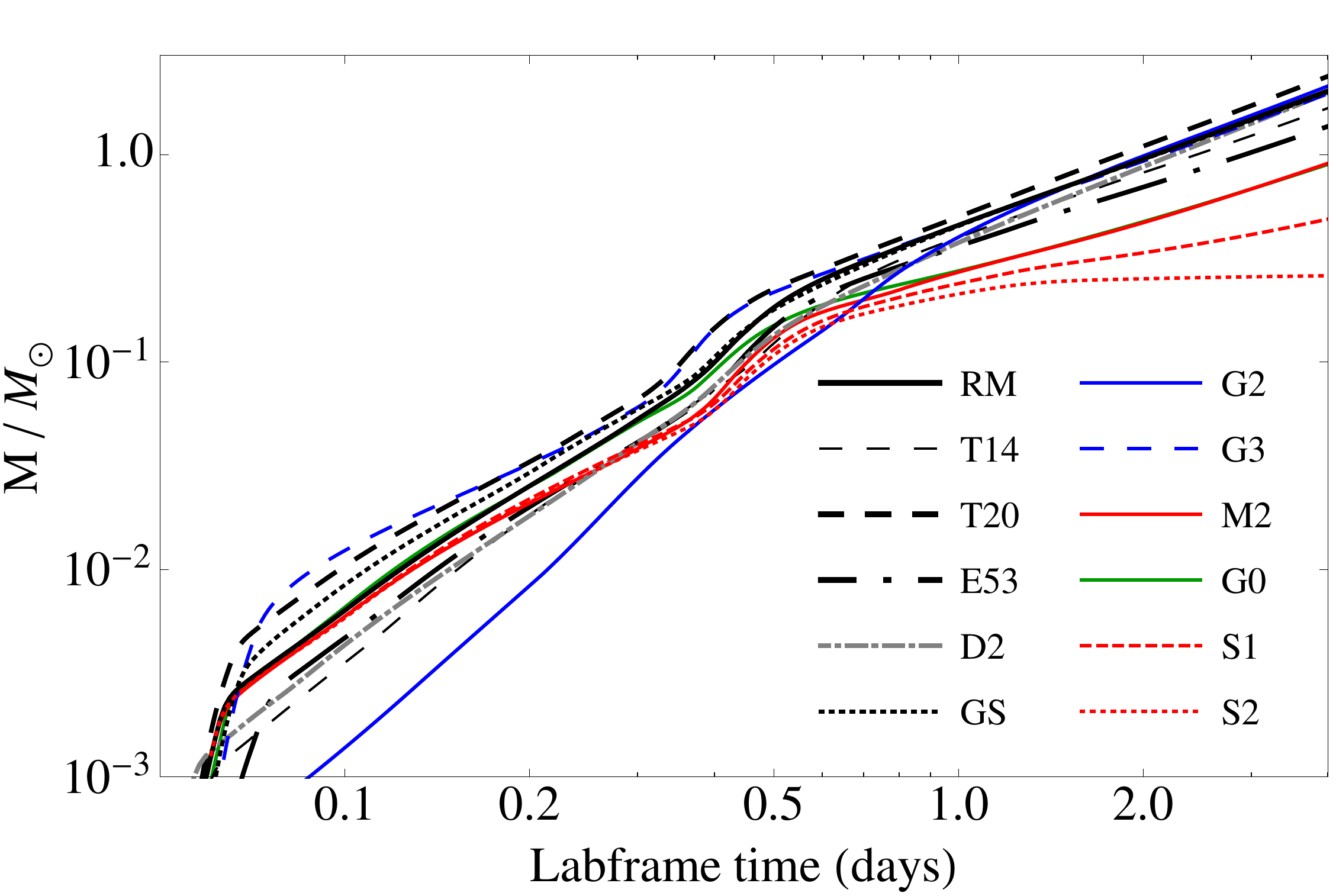}
\caption{Time evolution (in the laboratory frame) of the rest mass enclosed by the bubble blown by each of the models. The mass accounts for the contribution of the Northern and Southern hemispheres.}
\label{fig:hydroMass}
\end{figure}
\begin{figure}
\centering
\includegraphics[height=5.7cm]{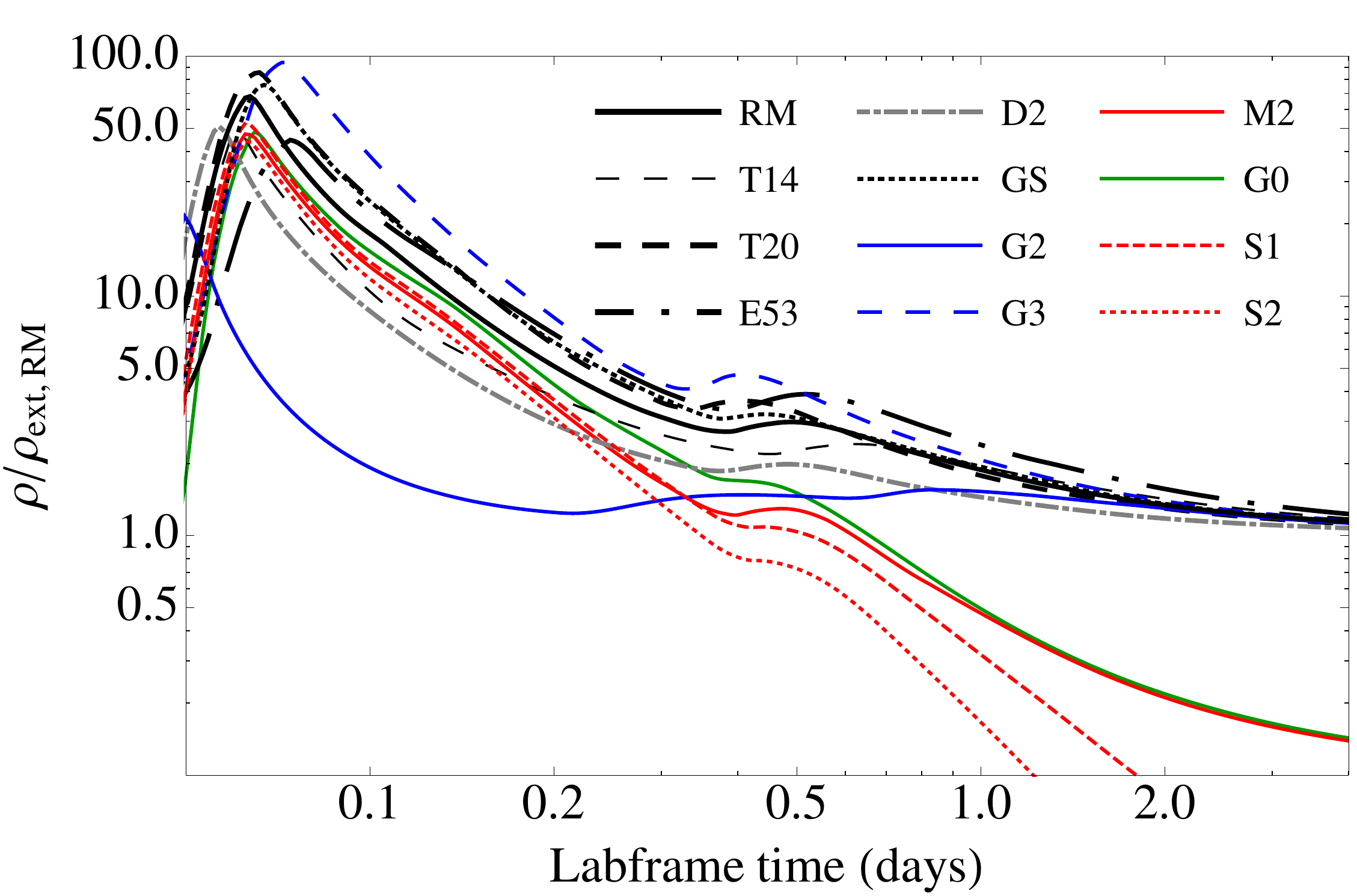}
\caption{Time evolution (in the laboratory frame) of the average bubble density of the bubble blown by each of the models in units of the EM density of the RM ($\rho_{\rm ext, RM}=8\times 10^{-14}\,$g\,cm$^{-3}$).}
\label{fig:hydroDensity}
\end{figure}
\begin{figure}
\centering
\includegraphics[height=5.7cm]{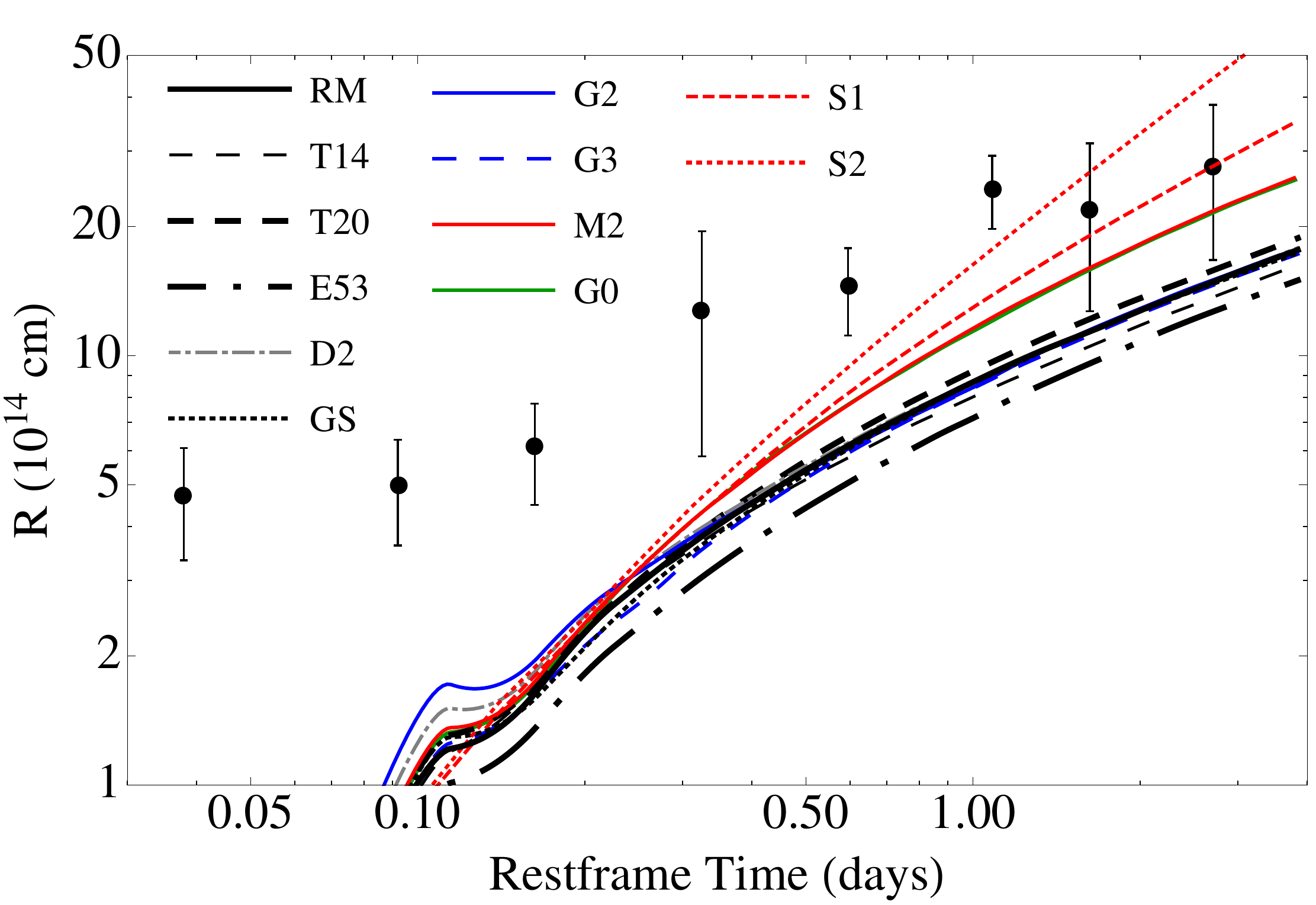}
\caption{Time evolution (in the rest frame of the source) of the transversal radius of different numerical models. The circles with error bars display the data obtained in T11 from fits to the observed flux assuming a simple BB spherical expansion of an emitting source. To convert the values of the radius to the rest frame, we employ a redshift $z=0.847$. Model G0 lies almost exactly below model M2.}
\label{fig:hydroRadius}
\end{figure}

In both models (RM and E53) the mass of the CE shell is $0.26 M_\odot$, but the total bubble mass exceeds that value by $\simeq 1\,$d (Fig.~\ref{fig:hydroMass}).%
\footnote{We note that the time axis in Figs~\ref{fig:hydroMass}~and~\ref{fig:hydroDensity}  corresponds to the lab-frame time, i.e. the time of an observer attached to the source. This time should not be confused with the `rest-frame' observer time shown in, e.g., Fig.~\ref{fig:hydroRadius} which is the time, $t_{\rm det}$, measured by a distant observer. That observer receives the information from the source by means of photons, but is assumed to be sufficiently close so that cosmological effects are unimportant. See the precise definition of $t_{\rm det}$ in Section~\ref{sec:originthermal}.} Only a fraction of the rest mass of the bubble comes from the matter dragged from the jet/CE-shell interaction region during the early phases of the evolution. Indeed, all models end up fully ablating the CE shell, which is incorporated into the bubble. A major fraction of the mass, however, comes from the EM which is swept up by the external shock and piles along the bubble surface. The mass enclosed by the bubble grows with the energy of the jet. More energetic models expand faster and, as a consequence, sweep matter of the EM most rapidly. After about 0.5\,d, the rate of mass growth decreases. This is the time at which a major fraction of the CE shell is ablated by the jet.

Because of its most rapid expansion, the average density of the bubble in RM is smaller than in E53 model (compare the dash-dotted and solid black lines in Fig.~\ref{fig:hydroDensity}). Thus, the average density of the bubble becomes smaller as we increase $E_{\rm iso}$.  We also note that the generic evolution of the average density displays a fast rise up to a maximum at times $\lesssim 2\,$h, and then a slower decrease. The time at which the maximum average density is reached increases as $E_{\rm iso}$ decreases. This behaviour is connected to the fact that the jet/CE-shell interaction is stronger initially, when the jet is more relativistic and is either still being injected at constant rate or decaying as $t^{-5/3}$ (note that $T_1=1100\,$s and $T_2=3800\,$s; Section~\ref{sec:jetparameters}). Hence, mass from the CE shell is quickly incorporated into the jet cocoon, causing the average bubble density to grow as well. Soon after the moment at which jet injection power is decreased ($T_2=3800\,$s), the rate of mass loading of the bubble from the CE shell decreases and produces the slow decline observed for $T \gtrsim 0.1\,$d.

We have also computed the time evolution of the cross-sectional radius of the bubble and found that both models display a similar transversal expansion if the EM is uniform. For reference in Fig.~\ref{fig:hydroRadius} we also display the evolution of the cross-sectional radius obtained from the simple model of T11, in which the observed flux is fit to an expanding BB with radius $R$ and effective temperature $T$. As we shall demonstrate in Section~\ref{sec:originthermal}, in our case most of the thermal emission is originated from a relatively small region compared with the cross-sectional radius of the bubble. In contrast, the results of T11 assume that the size of the emitting region is that of the expanding BB fit. Our models provide a typical size (estimated by its cross-sectional radius) which is similar to (but typically smaller than) that obtained with the (over) simplified physical model of T11 for the emitting region. 

\begin{figure*}
\centering
\includegraphics[height=8cm]{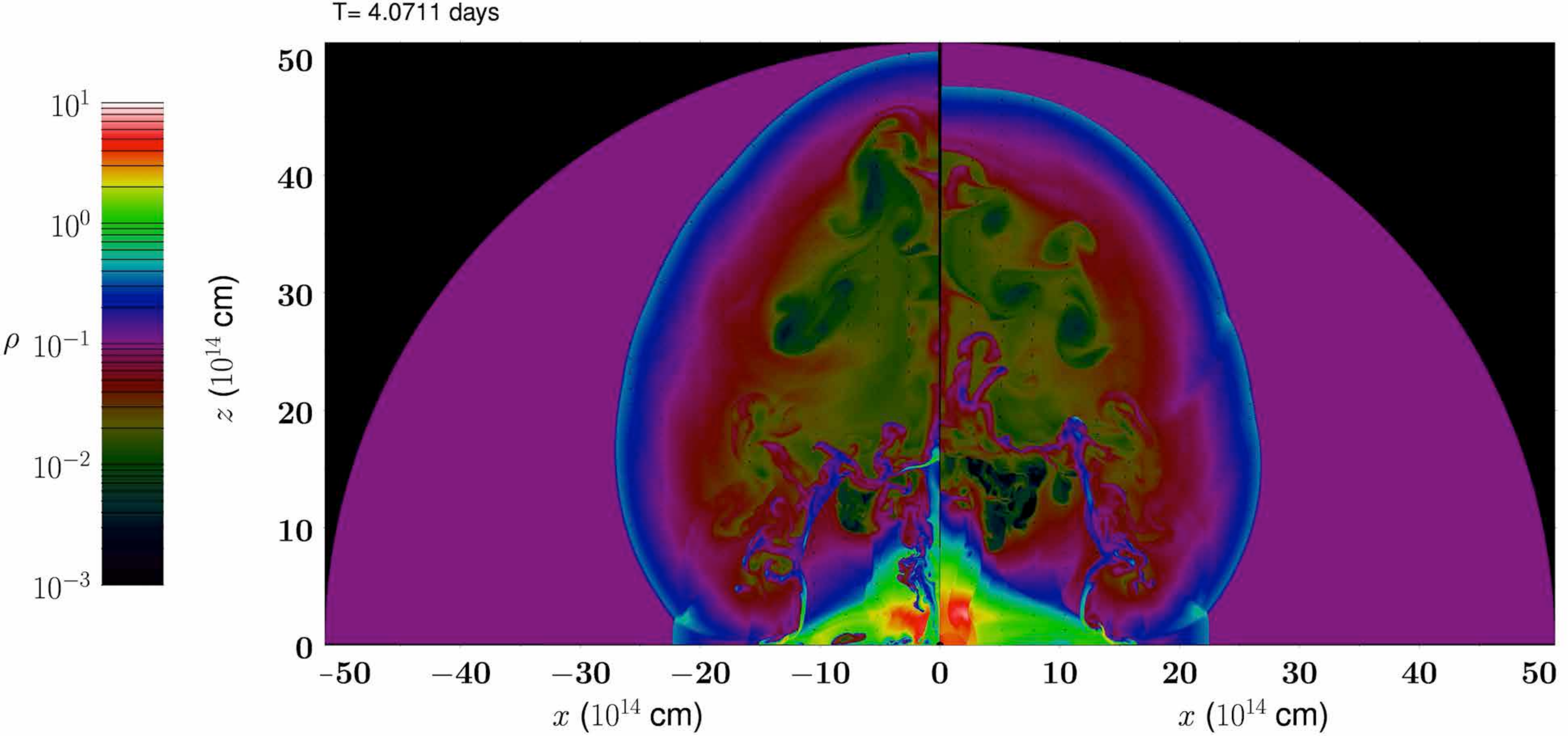}
\caption{Final distribution of the rest-mass density for the models with uniform, low-density EM ($\rho_{\rm ext} = 8 \times 10^{-15}$ g cm$^{-3}$). Left-hand panel: model M2. Right-hand panel: model G0 (same as M2 but without the `gap' between $R_{\rm 0}$ and $R_{\rm CE,in}$).}
\label{fig:hydroM2G0}
\end{figure*}

\subsubsection{Half-opening angle, $\theta_{\rm j}$}

Fixing the isotropic equivalent jet energy to the value in the RM, $E_{\rm iso} = 4 \times 10^{53}$\,erg, we have varied the jet half-opening angle, and considered  three cases for $\theta_{\rm j}$: $14^\circ$ (T14, Fig.~\ref{fig:hydro-models}d), $17^\circ$ (RM, Fig.~\ref{fig:hydro-models}a) and $20^\circ$ (T20, Fig.~\ref{fig:hydro-models}e). In all three cases, we set up the jet injection half-opening angles to be much wider than the innermost half-opening angle of the funnel $\theta_{\rm f,in}$ (see Fig.~\ref{fig:shellgeometry}). This is a basic ingredient of our model, since a very narrow jet would minimize the interaction with the CE shell, while an excessively broad jet, would be incompatible both with the theoretical expectations of the jet half-opening angle, and with the typical estimates based on observations connecting light-curve breaks with the jet angular size. The chosen range of values of the jet half-opening angle satisfies $\theta_{\rm j}\gg \theta_{\rm f,in}$.

 In order to understand how the variation of  $\theta_{\rm j}$ affects the dynamics, we first note that the true jet energy, $E_{\rm j}$, depends on $E_{\rm iso}$ and $\theta_{\rm j}$ so that by changing the jet injection half-opening angle the true injected jet energy is modified, although the amount of energy per unit solid angle remains constant. We note that the true jet energy of model T20 ($E_{\rm j}({\rm T20})=1.21\times 10^{52}\,$erg) is the largest among of all our models. We can appreciate that bubbles in models with larger jet half-opening angles have a larger radius in both longitudinal and transversal direction, and a more oblate structure. This is a consequence of increasing the jet energy as $\theta_{\rm j}$ in increased, since then the jet/CE-shell interaction region is larger, and it results in more massive bubbles (Fig.~\ref{fig:hydroMass}). 

Compared with the evolution of jets with smaller opening angles we observe that the mass growth rate of model T20 is qualitatively similar to that of the T14 and the RM models, but the transition to a smaller mass growth rate happens earlier. Indeed, the smaller the jet half-opening angle, the later such transition happens (Fig.~\ref{fig:hydroMass}).  Coupled to this transition, we can see that the bump in the average rest-mass density of the bubble (around $0.4$--$0.7$\,d; Fig.~\ref{fig:hydroDensity}) happens later for smaller values of $\theta_{\rm j}$.

The cross-sectional radii of models with increasing jet half-opening angles are very similar (Fig.~\ref{fig:hydroRadius}). However, in the long term, jets with larger half-opening angles exhibit slightly larger cross-sectional radii. This is in large part due to the larger true jet energy of models with larger $\theta_{\rm j}$.
\begin{figure*}
\centering
\includegraphics[height=8cm]{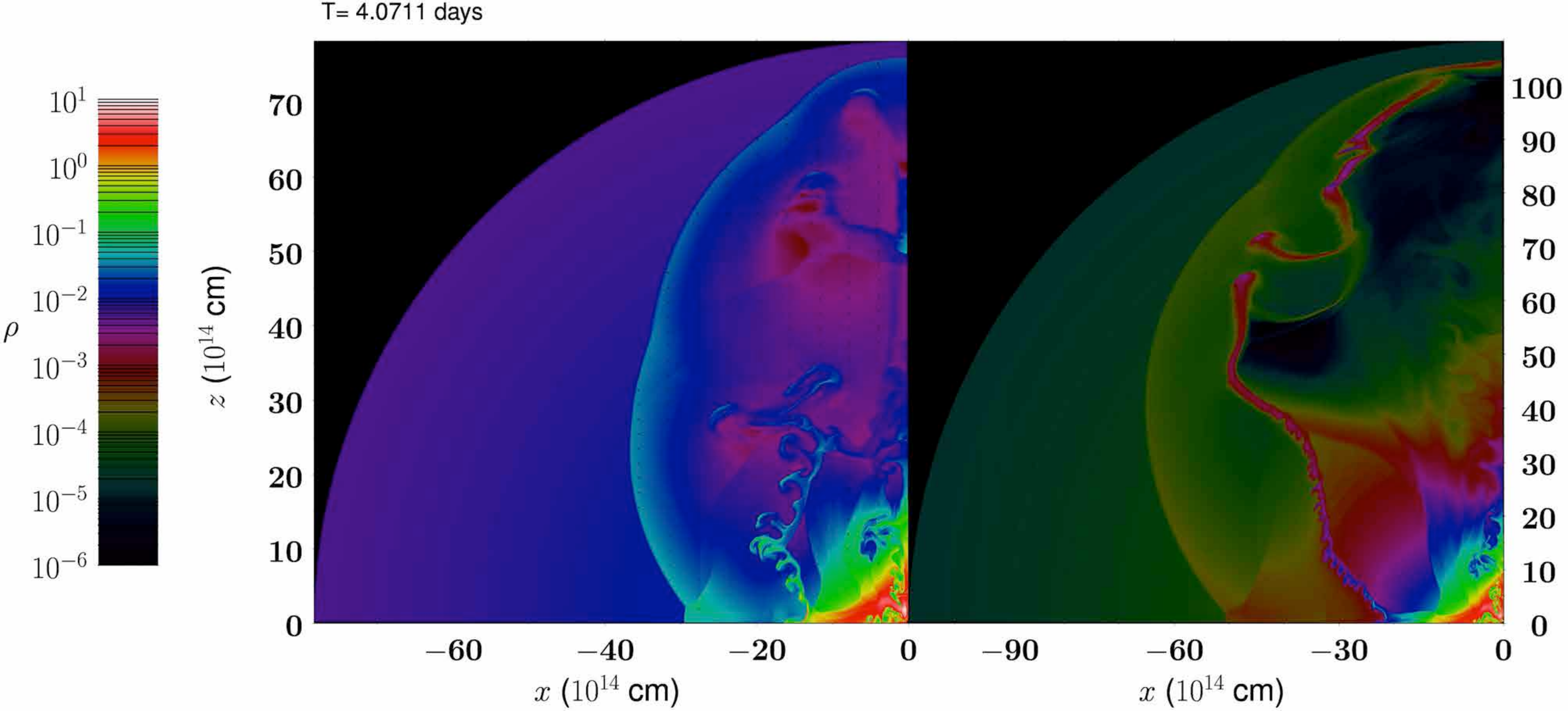}
\caption{Final distribution of the rest-mass density for the models with stratified EM. Left- (right)-hand panel: model S1 (S2), with a radial distribution of rest-mass density and pressure proportional to $r^{-1}$ ($r^{-2}$). Note the difference in the $z$-scales shown to the left and to the right of the corresponding panels, and the difference with respect to Fig.~\ref{fig:hydro-models} in the range displayed by the colour palette.}
\label{fig:hydroStra}
\end{figure*}

\subsubsection{CE-shell density contrast with respect to the external medium, $\rho_{\rm CE,sh}/\rho_{\rm ext}$}
\label{sec:CEshelldensity}

The ratio of rest-mass density of the CE shell to the EM density also plays an important role shaping the dynamics. We have tested three different CE-shell densities, $\rho_{\rm CE,sh}/\rho_{\rm ext} = 1500$ (RM, Fig.~\ref{fig:hydro-models}a), $817$ (D2, Fig.~\ref{fig:hydro-models}f) and $15000$  (D3, Fig.~\ref{fig:hydro-models}c) corresponding to masses of $M_{\rm CE,sh} \sim 0.26$, $0.14$, and $2.6 M_\odot$, respectively. As we shall see in Section~\ref{sec:originthermal}, the model with the most massive CE shell (D3),  yields a thermal signature incompatible with the observations and, thus, we will not consider it in this section for further discussion.

In the D2 model, the bubble is less dense than the RM and also less dense than most of the rest of the models in this parametric study (Fig.~\ref{fig:hydroDensity}). Since the volume of the bubble blown by the jet of model D2 is quite similar to that of the RM, its bubble mass is also smaller  (Fig.~\ref{fig:hydroMass}).  As we have seen in the RM, the mass of the bubble includes that of the swept EM as well as that accumulated  during the CE-shell/jet interaction. The former contribution is roughly similar in the RM and in the D2 model. However, the contribution to the bubble mass from the shell is substantially smaller, because of the lower CE-shell rest-mass density. In Fig.~\ref{fig:hydro-models}(b), we can observe that the RM displays a larger density in the central part of the bubble (close to the axis and to the origin, extending for about $10^{15}\,$cm).

Since the mass incorporated from the CE shell is smaller in model D2, it initially ($t\lesssim 0.3\,$d) expands faster than the RM (Fig.~\ref{fig:hydroRadius}). After that, the cross-sectional radius evolution is dominated by the mass incorporated to the cavity from the EM in relation to the energy supplied by the jet to the cavity (which is the same in both models) and, hence, the cross-sectional size of models D2 and RM become almost indistinguishable. 

For completeness, we have tested a simple stratification of the CE shell in which the rest-mass density and pressure decrease as $\propto r^{-2}$ (model GS, Fig.~\ref{fig:hydro-models}g). We have set the rest-mass density $\rho_{{\rm CE,sh},0}/\rho_{\rm ext} = 4304$ at $r = R_{\rm CE,in}$ in order to have, approximately, the same mass in the CE shell as in RM. The pressure at $r = R_{\rm CE,in}$ is the same as in the EM, i.e. $p_{{\rm CE,sh},0} = p_{\rm ext} = 10^{-5}c^2 \rho_{\rm ext}$. The global hydrodynamical properties like the bubble mass (Fig.~\ref{fig:hydroMass}), average cavity density (Fig.~\ref{fig:hydroDensity}), and cross-sectional radius (Fig.~\ref{fig:hydroRadius}) are very similar in this model to those of the RM. However, as we shall see in Section~\ref{sec:CEgeometry}, the stratification of the CE shell modifies jet/CE-shell interaction and imprints substantial changes in the computed thermal emission.

\subsubsection{External medium}
\label{sec:externalmedium}

In the previous sections, we have always considered a uniform EM (i.e. isopycnic and isobaric). However, the environment of massive stars is certainly more complicated than in our simple model (see Section~\ref{sec:progenitormodels}). Such complex environments can also have rather complicated density profiles. A suitable simplification that helps us disentangling the many different effects that show up in the dynamics of our jets is to consider first that the EM is a uniform medium. Later, we will parametrize the EM assuming that the rest-mass density decays as a power law of the distance.

The fiducial value of $\rho_{\rm ext}$ for our RM, yields an EM mass of $\sim 6M_\odot$ within the numerical domain. This value is a balance between what we realistically expect in the environment around the secondary star of the merger (likely, a lower value of the EM mass; Section~\ref{sec:progenitormodels}) and the numerical difficulty posed by the very large density jump between the CE shell and the EM, together with very low pressure-to-density ratios in the RM ($p_{\rm ext}/\rho_{\rm ext} = 10^{-5}c^2$). In order to assess the effects on the dynamics and on the light curves of a less dense environment we have reduced the density of the EM by one order of magnitude in models M2 and G0 ($\rho_{\rm ext} = 8 \times 10^{-15}$\,g\,cm$^{-3}$; Table~\ref{tab:params}). To easy the numerical difficulty of reducing the EM density, while keeping the same mass in the CE shell (which means that the density contrast $\rho_{\rm CE,sh}/\rho_{\rm ext}$ is 10 times larger in the M2 and G0 models than in the RM), we increase the ratio pressure-to-rest-mass density everywhere in the domain (i.e. for models M2 and G0 we have $p_{\rm ext}/\rho_{\rm ext} = 10^{-4}c^2$). The only difference between both models is that in G0 we have extended the inner radius of the CE shell until the innermost boundary at $R_0$. As we can see  in Fig.~\ref{fig:hydroM2G0}, the jets of models M2 and G0 are able to reach larger distances in the same evolutionary time than the jet of the RM, and the cavity blown by the jet in model M2 has a more prolate shape. The cavity of model G0 is a bit more spherical than that of M2 since the jet has to pull the extra initial mass where the `gap' was located. Thus, in order to properly compute the late light curves and spectra, we have extended the computational domain up to $R_{\rm f} = 5.13\times 10^{15}$\,cm, so that the EM in models M2 and G0 encloses a mass of $2.3 M_\odot$. We note that up to the same outer boundary than in the RM, the mass of the EM of both models is only of $\sim 0.6M_\odot$. 

The evolution of mass ploughed by the bubble (red solid line in Fig.~\ref{fig:hydroMass}) is similar to that of RM until $0.5$\,d, where the evolution is still dominated by the interaction with the CE shell. The average rest-mass density (Fig.~\ref{fig:hydroDensity}) also displays a similar behaviour until this time and taking similar values as that RM. At longer times the average density becomes obviously smaller by a factor of $\sim 10$, as expected because of the 10 times smaller EM density of model M2 with respect to the RM. The cavity blown by the jet in the M2 model is not only longer, but also has a factor of $\sim 2$ larger transversal radius than the RM (Fig. \ref{fig:hydroRadius}). As a concluding remark in the case of models with uniform EM, and as it is expected, a smaller EM rest mass modifies the long-term evolution (i.e. the evolution after about 0.5\,d), but the initial CE-shell/jet interaction does not change appreciably. We advance that a uniform medium cannot be extended to arbitrarily large distances from the progenitor since it would bring an unrealistically large mass in the EM.

We now turn to models with a non-uniform EM and consider two simple parameterizations of stratified external environments. We assume that the rest-mass density and pressure decrease with the distance as $r^{-1}$ (S1 model) or $r^{-2}$ (S2 model) from $r=R_{\rm CE,in}$. Below the CE shell, i.e. in the region $R_0<r < R_{\rm CE,in}$ we impose a uniform medium with the same rest-mass density as the in the uniform ambient medium models, but with a larger pressure $p_{\rm ext}/\rho_{\rm ext}=10^{-3}c^2$. The pressure in the CE shell is the same as in the RM. We note that, differently from \cite{DeColle_etal_2012ApJ...751...57}, with this initialization, any potential jet break (which would occur if the CE shell was absent), would happen at very different distances depending on the rest-mass density gradient. 
\begin{figure}
\centering
\includegraphics[height=5.4cm]{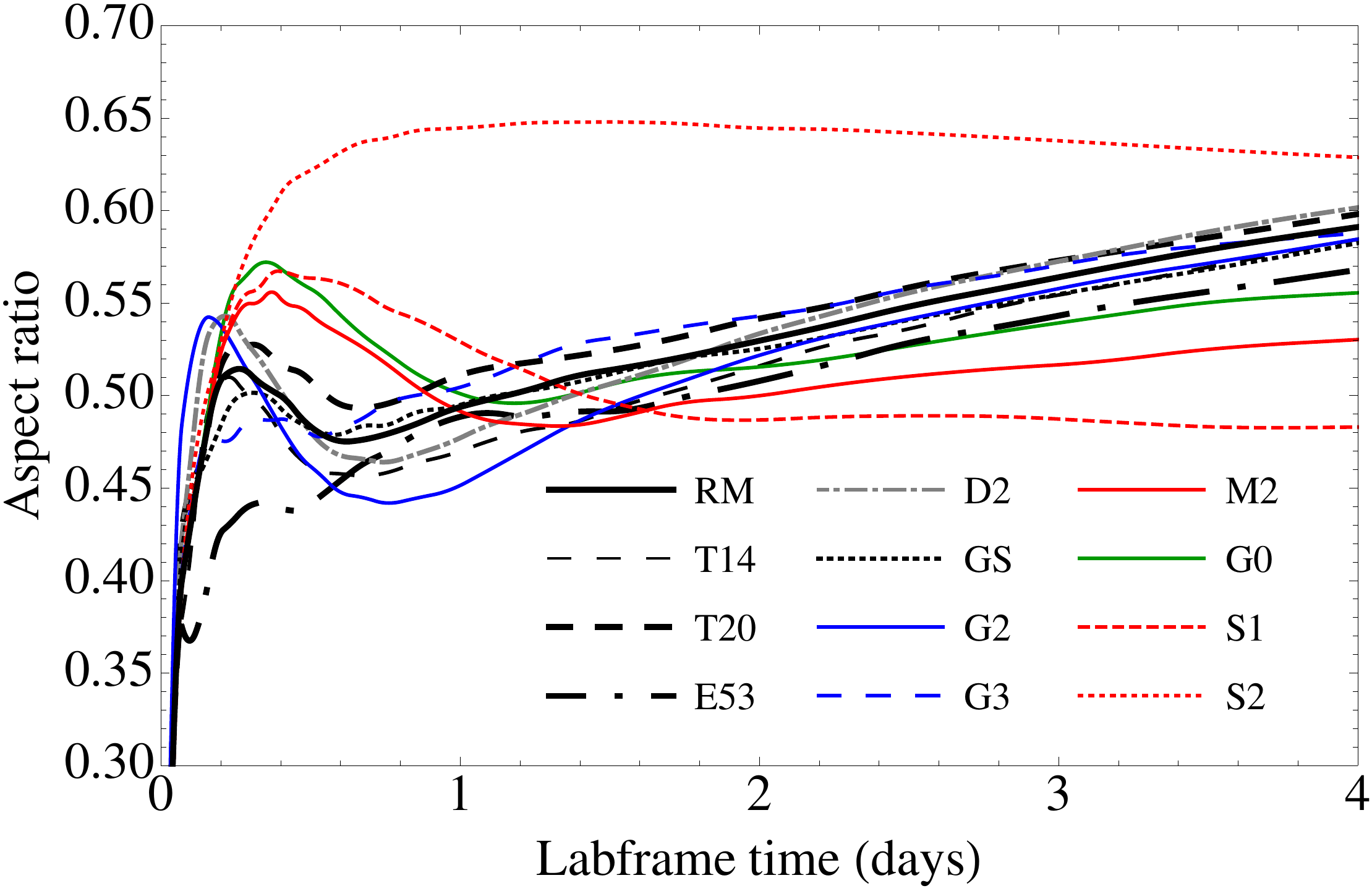}
\caption{Evolution of the aspect ratio of different models in the laboratory time. This aspect ratio is defined as the ratio between the cross-sectional diameter and the longitudinal (along the $z$-axis and including the Northern and Southern hemispheres) extent of the jet.}
\label{fig:aspect-ratio}
\end{figure}

Jets propagating in wind-like media tend to develop more elongated cavities than the same jets moving through a uniform, high-density medium in the long term (Fig.~\ref{fig:hydroStra}). Likewise, the sideways expansion is also larger compared with jets propagating in a isopycnic/isobaric medium (Fig.~\ref{fig:hydroRadius}). Since the interaction at early times is determined mostly by the conditions in the CE shell, the average rest-mass density and total mass of the jet approaches the RM values until $\sim 0.4\,$d. But, as the jet proceeds through the stratified medium, both quantities differ at later times (see Figs~\ref{fig:hydroMass} and \ref{fig:hydroDensity}). The mass of the bubble grows slowly and its average density, which decreases as $t^{-2}$ in the S1 model and as $t^{-3}$ in the S2 model, is also smaller compared with the RM.

\subsection{Long-term evolution}

In the long term, all models moving in a uniform, high-density ambient medium develop a quasi-spherical cavity (Fig.~\ref{fig:hydro-models}).  However, a simple (spherically symmetric) Blandford--McKee blastwave is not adequate to describe the dynamics during the first days of evolution, and the reason is the jet/shell interaction. Since both the EM and the CE shell are much denser than the jet, it develops a mildly relativistic bow (forward) shock and a non-relativistic reverse shock. The jet/shell interaction causes the jet to decelerate and produce a `hot bubble' in which the original jet is disrupted. If this kind of scenario would yield a typical afterglow, no signs of jet break would have been observed, since there is no jet anymore. 

After an initial phase dominated by the CE-shell/jet interaction dynamics (lasting for $\sim 1\,$d), the cross-sectional diameter of the bubble expands faster than the longitudinal jet dimension. This happens when the cavities travel a distance of the order of the Sedov length ($l_{\rm Sedov}=\left( \frac{(17-4k)E_{\rm j}}{8\pi\rho_{\rm ext}c^2}\right)^{1/(3-k)}$, $k$ being the index of the power-law decay of the rest-mass density) for each model.\footnote{For the RM and most of our models endowed with a uniform, high-density medium the Sedov length is $\sim 4\times10^{14}\,$cm.} In Fig.~\ref{fig:aspect-ratio}, we quantify the aspect ratio of each model, defined as the ratio between the cross-sectional diameter and the longitudinal (along the $z$-axis) jet length. After $\sim 1$ d and until $4$ d the aspect ratio grows, becoming $\simeq 0.5$--$0.6$. Extrapolating the rate of increase of the aspect ratio between 1 and 4 d, we estimate that our models propagating in a uniform, high-density circumburst medium will become spherical (aspect ratio equals one) in approximately 12 to 18 d. This estimate is rather robust, since models enter a quasi-self-similar regime after $\simeq 1\,$d. The rate of growth of the aspect ratio $\simeq 0.03\,$units\,d$^{-1}$ is a generic feature, only weakly dependent on the jet parameters and properties of the CE shell. Thus, we find that this transition to sphericity roughly coincides with the time at which T11 find that a SN contribution is needed to explain the flattening of the light curves in the optical bands. Since the EM is less massive in models M2 and G0, the transition to sphericity is delayed and the shape of the blown cavity displays a slightly smaller aspect ratio than in other uniform models with larger EM rest-mass density.
\begin{figure}
\centering
\centering
\includegraphics[width=8.4cm]{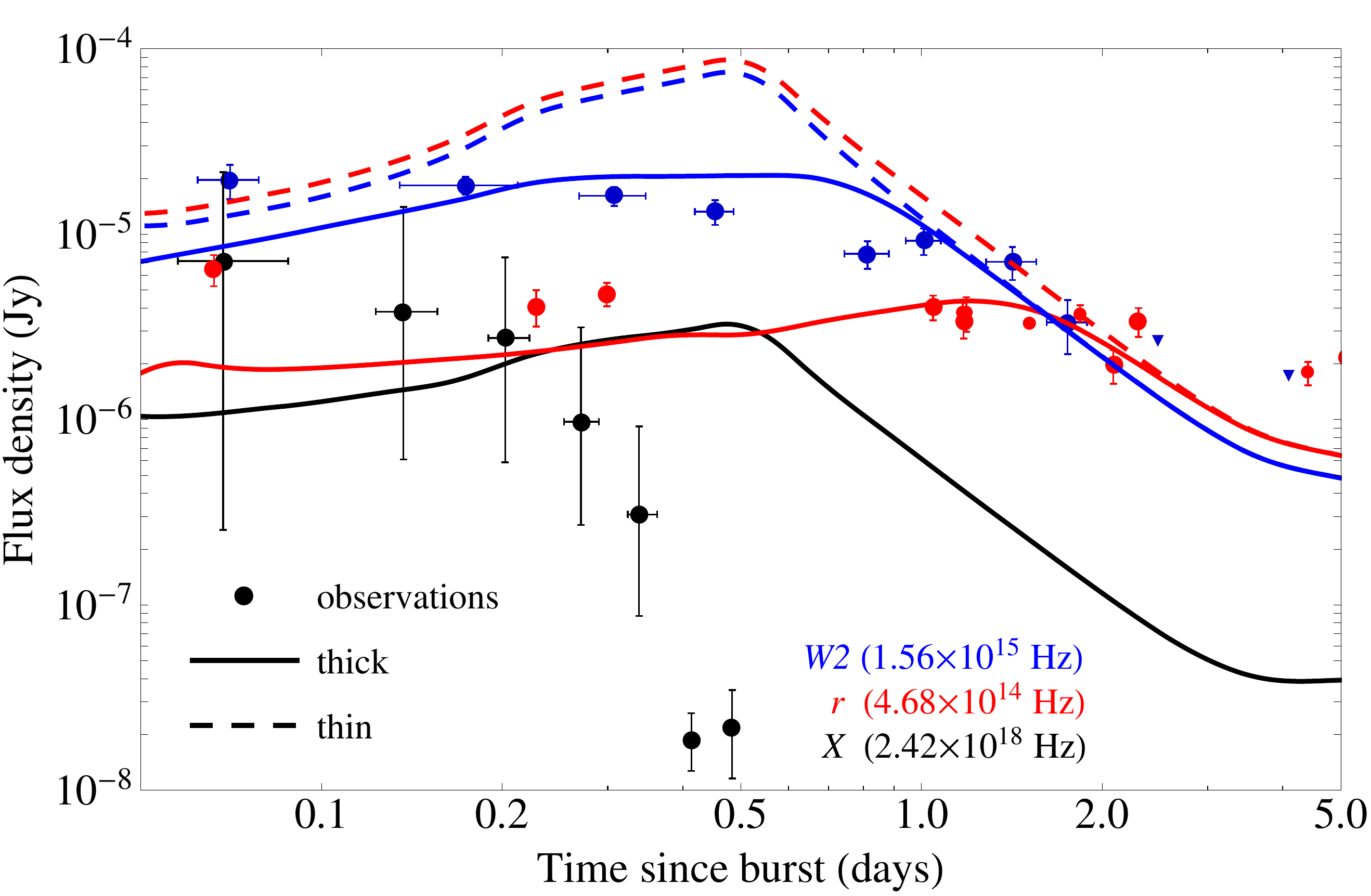}
\caption{Light curves for the RM considering only the (thermal) bremsstrahlung-BB contribution. Both optically thick (solid lines) and thin (dashed lines) light curves are plotted, to better illustrate the transition from optically thin to optically thick emission. For the X-ray band (black lines), the optically thin and thick light curves coincide, since the X-ray emitting region is optically thin. For the representation of the X-ray data, we have clustered the data of each of the XRT observing cycles into a single point, with error bars showing the data dispersion.}
\label{fig:referencethickthin}
\end{figure}

In contrast to the long-term evolution of jets propagating in a uniform medium, jets travelling along a stratified EM tend to develop prolate cavities. This feature is reflected in the decrease after 0.5\,d of the aspect ratio (Fig.~\ref{fig:aspect-ratio}). Eventually, the aspect ratio tends to settle to a roughly uniform value, since the jet encounters less resistance in all directions as it expands across the EM. We also note that the S2 model experiences a rapid transversal expansion after 1 d associated with the rapid initial decrease of rest-mass density and pressure in the EM. In case the evolution could be extrapolated forward in time, these cavities will take much longer time to become spherical. 

\section{Origin of the thermal emission}
\label{sec:originthermal}

In this section, we compute the thermal signature of a number of models (and specifically of the RM) with the goal of uncovering the provenance of the thermal emission. To do so, we post-process the output of our RHD simulations (using \tiny{MRGENESIS}\normalsize) with our radiative transport code \tiny{SPEV}\normalsize. Using \tiny{SPEV }\normalsize we can produce light curves and spectra accounting for or neglecting the absorption processes. We will refer to these two modes of computing the spectral properties of our models as `thick' or `thin', respectively. Comparing the thin and thick spectral properties we are able to better understand when the systems at hand become optically thin and where the emission and absorption dominantly take place. 
\begin{figure*}
\centering
\centering
\includegraphics[width=17.6cm]{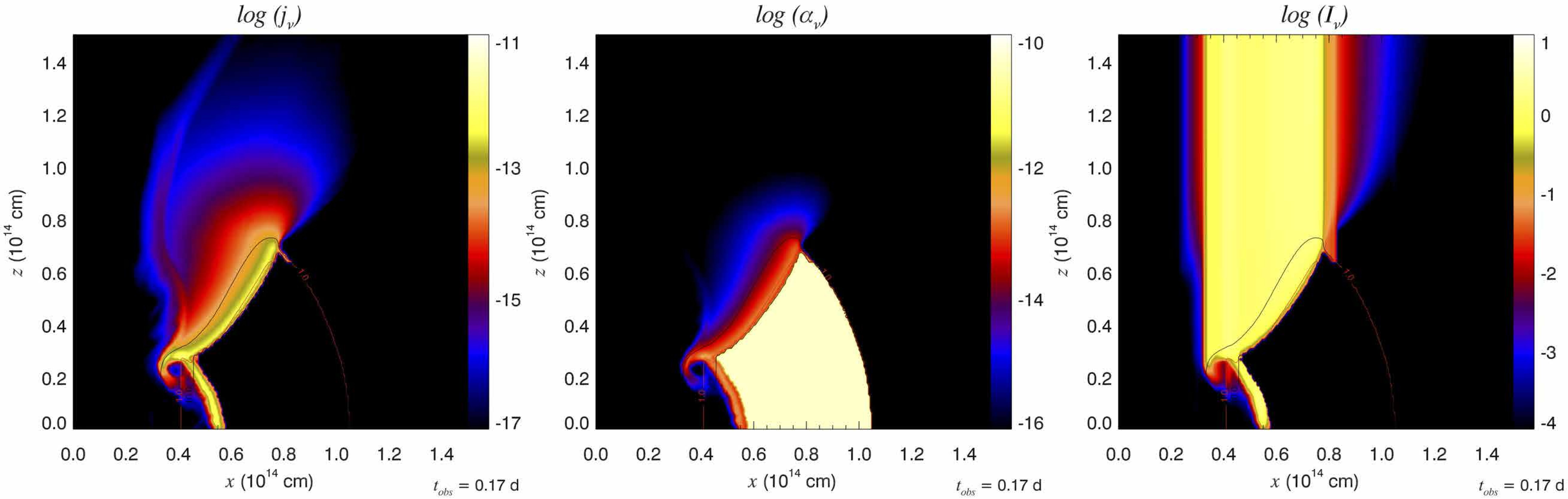}
\caption{Emission, $j_\nu$, (left) and absorption, $\alpha_\nu$, (centre) coefficients and evolution of the specific intensity, $I_\nu$ (right) along the line of sight. The observer is located in the vertical direction (towards the top of the page) at a viewing angle $\theta_{\rm obs} = 0^\circ$. The emission is computed in the $W2$ band for band free--free (thermal) bremsstrahlung process, at an observational time $t_{\rm obs} = 0.17$\,d. The units of $j_\nu$,  $\alpha_\nu$ and $I_\nu$ are given the CGS system (see Appendix~\ref{sec:spev} for details). From the figures, one can realize that the main contribution of the thermal radiation comes from the interaction region jet/CE-shell, located at a distance from the symmetry axis of $\simeq 3\times 10^{13}\,$cm and extending to $\simeq 8\times 10^{13}\,$cm. This emission region coincides with the locus of the section of the CE shell shocked by the relativistic jet.}
\label{fig:em-ab-in-obs}
\end{figure*}
\begin{figure*}
\centering
\centering
\includegraphics[width=17.6cm]{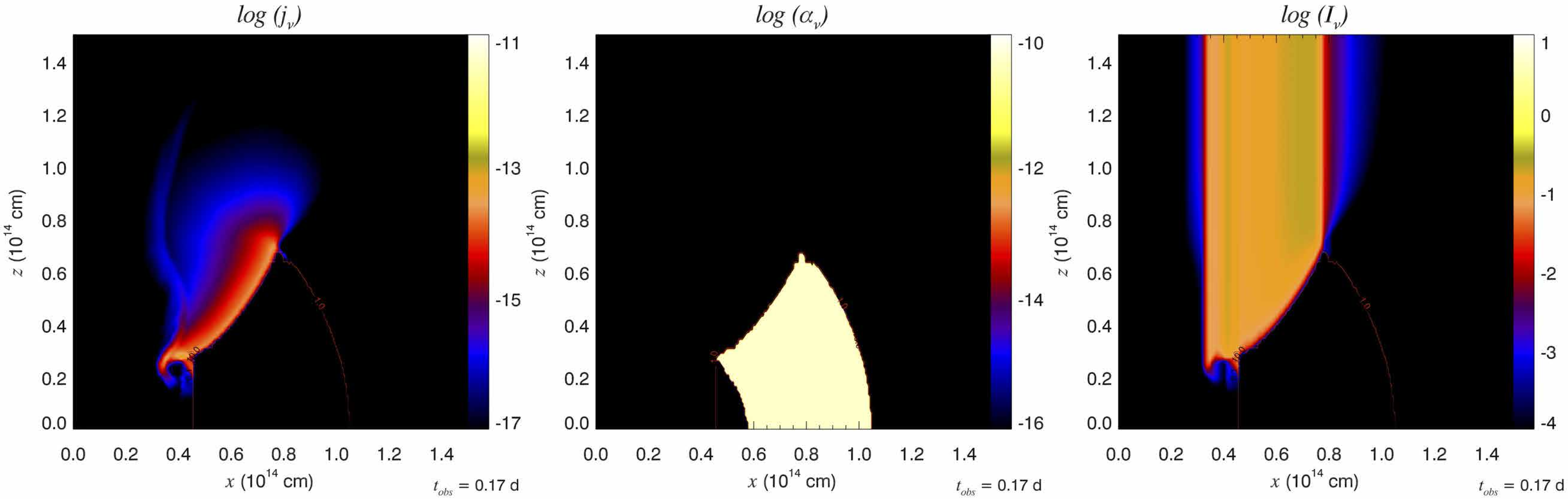}
\caption{Same as Fig.~\ref{fig:em-ab-in-obs} but in the X-ray band.}
\label{fig:em-ab-in-obs-X}
\end{figure*}
\begin{figure*}
\centering
\centering
\includegraphics[width=17.6cm]{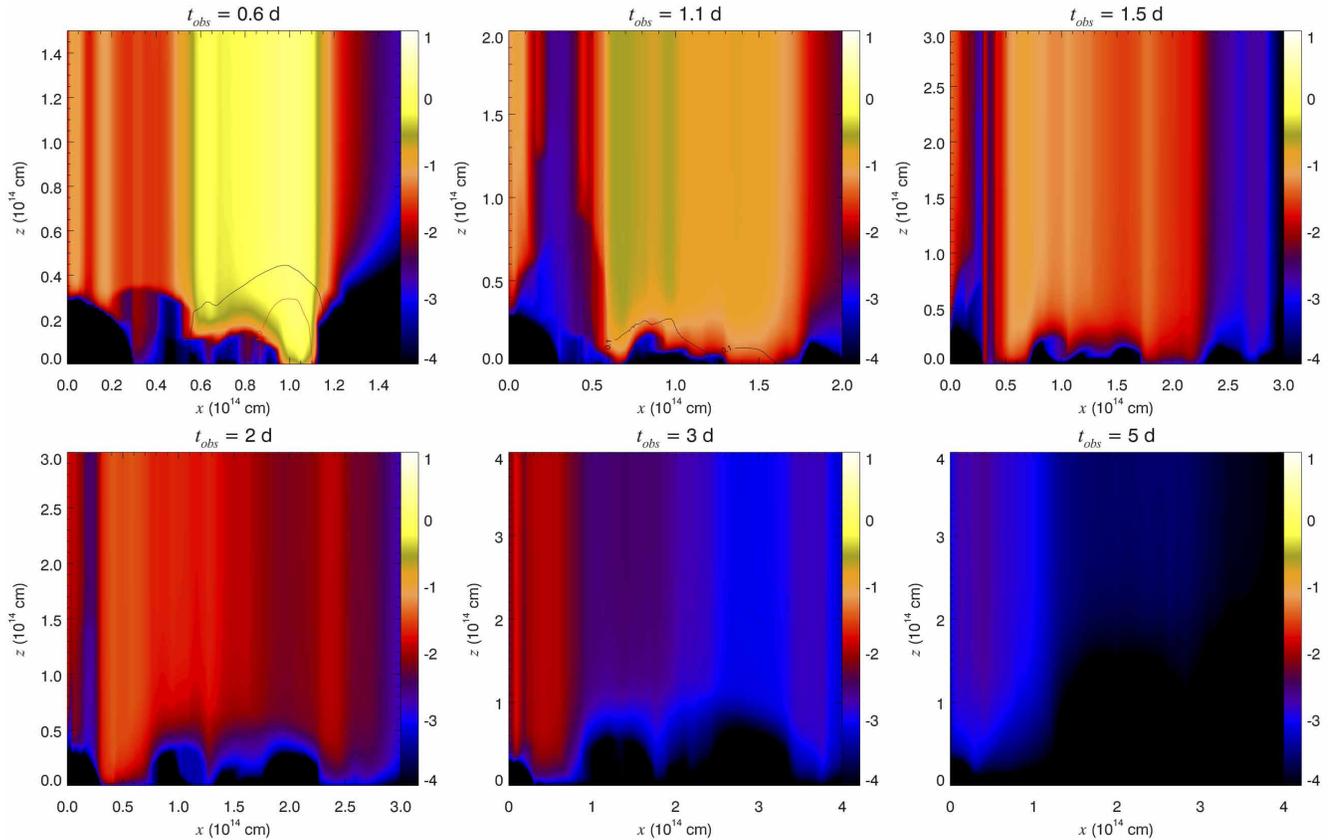}
\caption{Evolution of the specific intensity, $I_\nu$, in the $W2$ band (same as the right-hand panel in Fig.~\ref{fig:em-ab-in-obs}). The image is focused on the jet/CE-shell interaction region. Note that the transition from optically thick to optically thin at $\sim1.5$--$2$\,d (top-right and bottom-left panels) is due to the ablation of the CE shell, which is absent after $\sim 2\,$d (bottom panels). The observational times are provided above of each of the panels.}
\label{fig:in-obs-ev}
\end{figure*}

For the RM (Fig.~\ref{fig:referencethickthin}), we observe that the system is optically thick until about $\sim1$ d after the burst\footnote{Note that if the dashed and continuous blue lines overlap in Fig.~\ref{fig:referencethickthin} it means that absorption does not influence the observed emission, i.e. the medium has become optically thin.} in the $W2$ band, and until $\sim 2$ d in the $r$ band. In the X-ray band, the system is optically thin from the beginning of the observing time (note that the dashed black line overlaps with the solid black line in Fig.~\ref{fig:referencethickthin}). It is evident that the computed thermal emission in the X-ray band peaks too late (at about $0.5\,$d) compared with the observational data, though the flux decline after the emission peak happens at a rate compatible with the observed data. As we will show (Section~\ref{sec:CEgeometry}), these two facts are connected to a large extent to the {\em assumed} geometry and rest-mass distribution of the CE shell. Since the goal of this work is not obtaining a {\em perfect} fit of the data but understanding the basic properties of the system, we have not tuned the geometry of the channel to accurately describe the observations. Instead, we point out the qualitative fact that the time at which we find the maximum flux density depends on frequency: the larger the frequency the earlier the flux density peak happens. 

When the system becomes optically thin at all optical frequencies after $\sim 1.5$--$2$\,d, the thermal spectrum is inverted and we observe a larger flux in the $r$ band than in the $W2$ band, i.e. the initially blue system becomes red as the observations in T11 suggest. This feature is related to the time by which the CE shell is fully ablated by the ultrarelativistic jet. To demonstrate this assessment, we have identified the location of the parts of the system from where thermal radiation is coming from. This is not a trivial task, since in our method, the contribution to the total flux of each computational cell can be strongly \emph{blurred} because of the relativistic effects (e.g., time dilation, time delays, aberration). That means that for a given observed time, $t_{\rm obs}$, there will be contributions from different snapshots of the hydrodynamical evolution. We consider a virtual detector consisting of a screen oriented perpendicularly to the symmetry axis (i.e. at an observing angle of $0^\circ$). For a given laboratory time in our hydrodynamical simulations, $T$, the photons coming from a fluid element located at a distance $R$ (measured along the symmetry axis from the centre of the system) will arrive to the detector in a time $t_{\rm det} = T - R/c+T_{\rm offs}$, where $T_{\rm offs} = R_0 / c \sqrt{1-\Gamma_{\rm i}^{-2}}$ is defined as the (laboratory frame) time spent by the jet to travel from $r=0$ to $r=R_0$. The relation between $t_{\rm det}$ and the observer's frame time is given by $t_{\rm obs} = t_{\rm det} (1 + z)$, where here $z$ refers to the redshift. In Fig.~\ref{fig:em-ab-in-obs} (left-hand and central panels), we depict the emission and absorption coefficients of free--free bremsstrahlung process at $t_{\rm obs} = 0.17$\,d. We also show the specific intensity an observer looking head on the jet would see. We note that the dominant contribution to the flux accumulated in our virtual detector is due to the region where the jet has interacted more strongly with the CE shell, namely, in regions which extend from $0.3$ to $0.7\times10^{14}$\,cm in the $z$-direction and from $0.4$ to $0.75\times10^{14}$\,cm in the $x$-direction. It is clear that the CE shell is not emitting but absorbing all the flux coming from regions with $z < 0.3\times10^{14}$\,cm. 
\begin{figure}
\centering
\centering
\includegraphics[width=8.4cm]{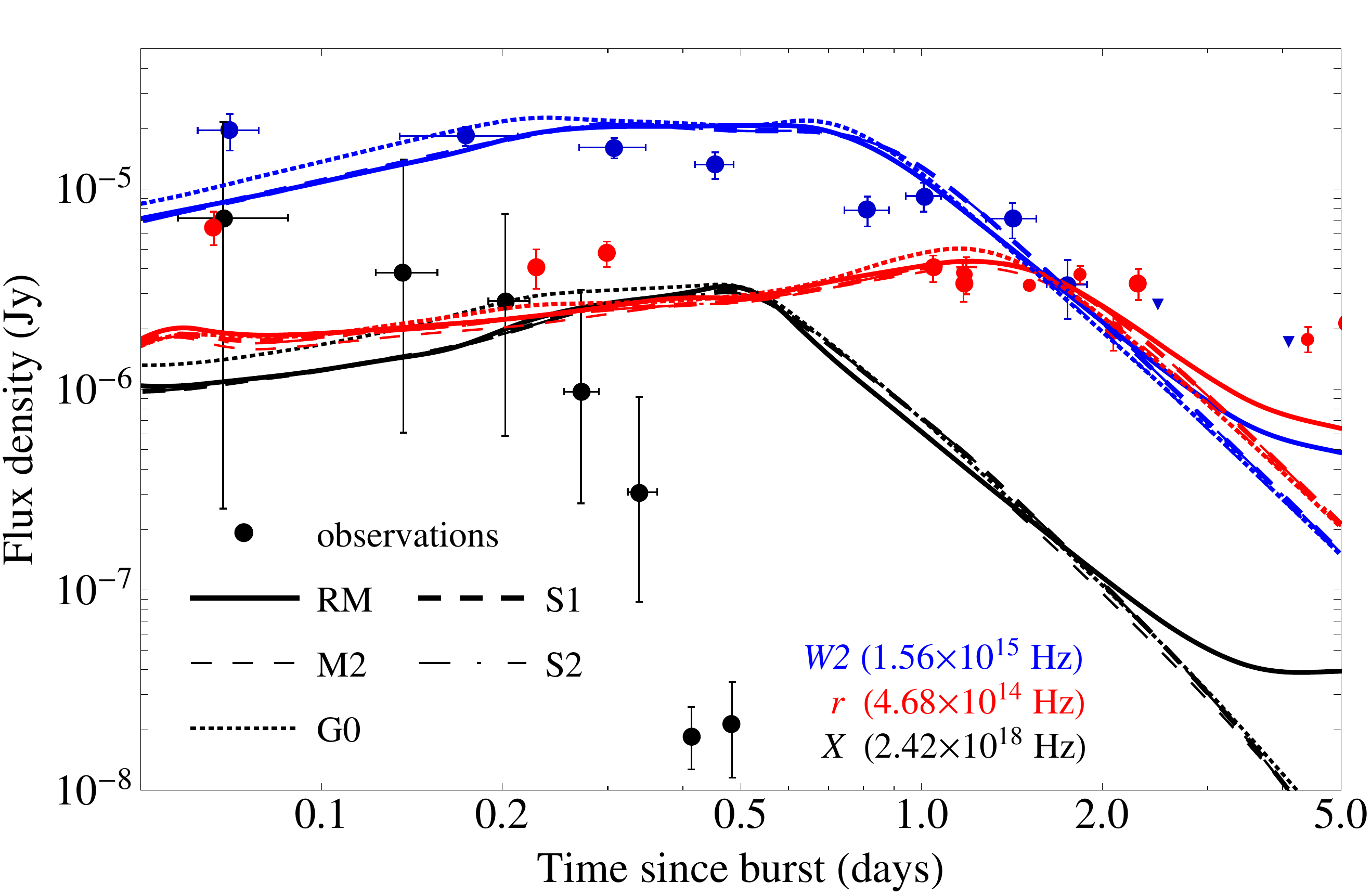}
\caption{Light curves for the RM (solid lines), M2 (thin dashed lines), S1 (thick dashed lines), S2 (dot--dashed lines) and G0 (dotted lines) considering only the (thermal) bremsstrahlung-BB contribution. Optically thick light curves are plotted.}
\label{fig:reference-vs-stra}
\end{figure}
\begin{figure}
\centering
\centering
\includegraphics[width=8.4cm]{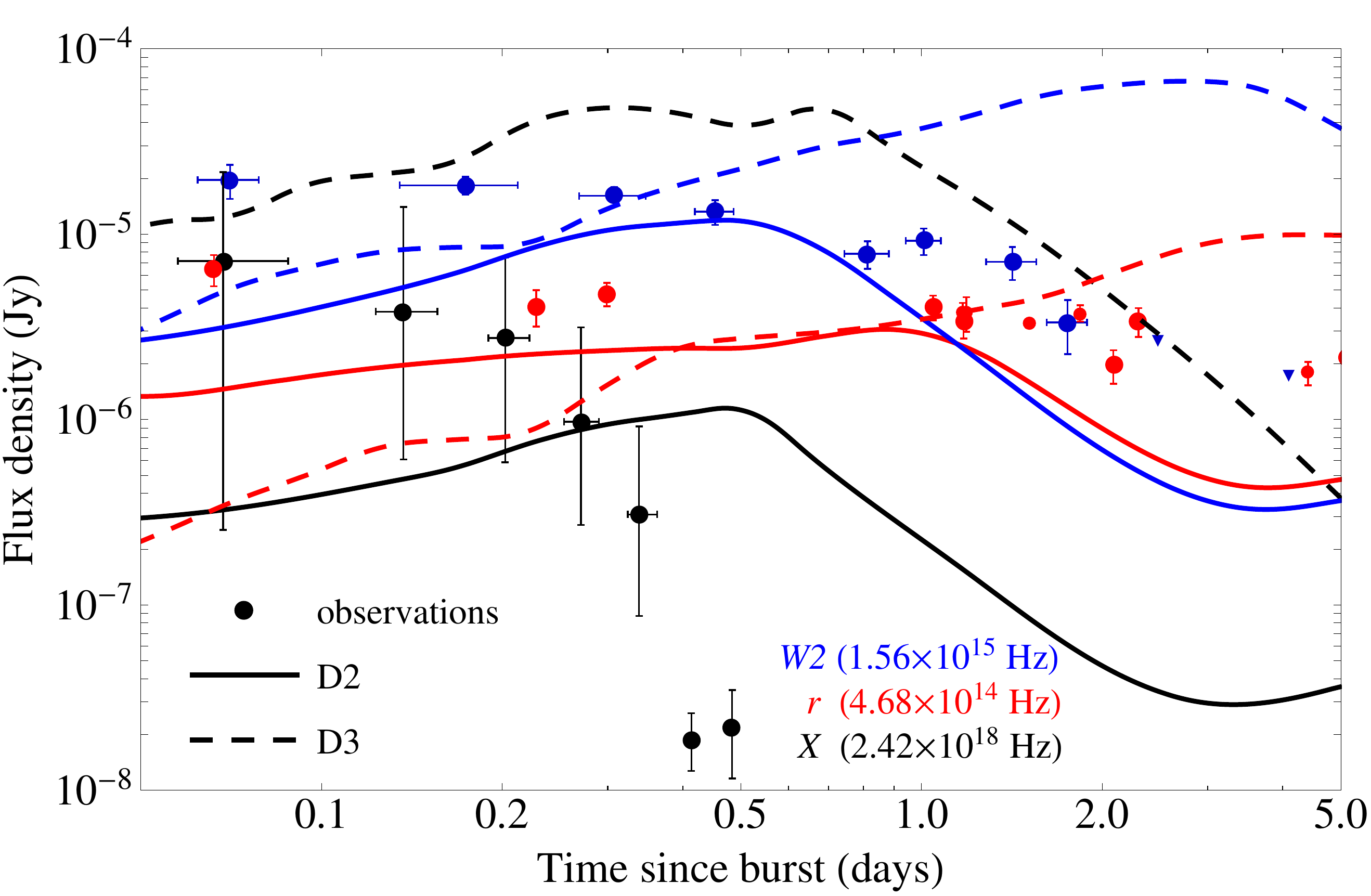}
\caption{Light curves for D2 (solid lines) and D3 (dashed lines) models, considering only the (thermal) bremsstrahlung-BB contribution. Optically thick light curves are plotted.}
\label{fig:D2D3models}
\end{figure}

From the spatial distribution of the specific intensity in the X-ray band (Fig.~\ref{fig:em-ab-in-obs-X}, right-hand panel) and the distribution of the emissivity (Fig.~\ref{fig:em-ab-in-obs-X}, left-hand panel) we conclude that the X-ray detectable region is smaller than at optical frequencies. This region is concentrated very close to the surface of the CE shell facing the symmetry axis. The extent of the X-ray observable emitting region (facing up in Fig.~\ref{fig:em-ab-in-obs-X}, right-hand panel) is strongly dependent on the CE-shell geometry and mass distribution. A less dense shell closer to the symmetry axis would enhance the observed emission and, since this region would be dredged up by the jet faster than the current high-density CE shell, its emitted flux would decrease much sooner than in our models (see Section~\ref{sec:CEgeometry}). 

Figure~\ref{fig:in-obs-ev} displays several snapshots of the evolution of the specific intensity in the $W2$ band for different observer's frame times, showing the process of ablation of the shell and the consequent reduction of emission. We notice that the CE shell is almost complete at $0.17$\,d (Fig.~\ref{fig:em-ab-in-obs}), while it is strongly disrupted (almost ablated) at $0.6$\,d (Fig. \ref{fig:in-obs-ev}, upper-left panel). During the subsequent evolution the optical depth decreases drastically due to the ablation process suffered by the CE shell, yielding a transition from an optically thick to an optically thin regime, as well as triggering a reddening of the observed system.

\begin{figure*}
\centering
\centering
\includegraphics[width=17.6cm]{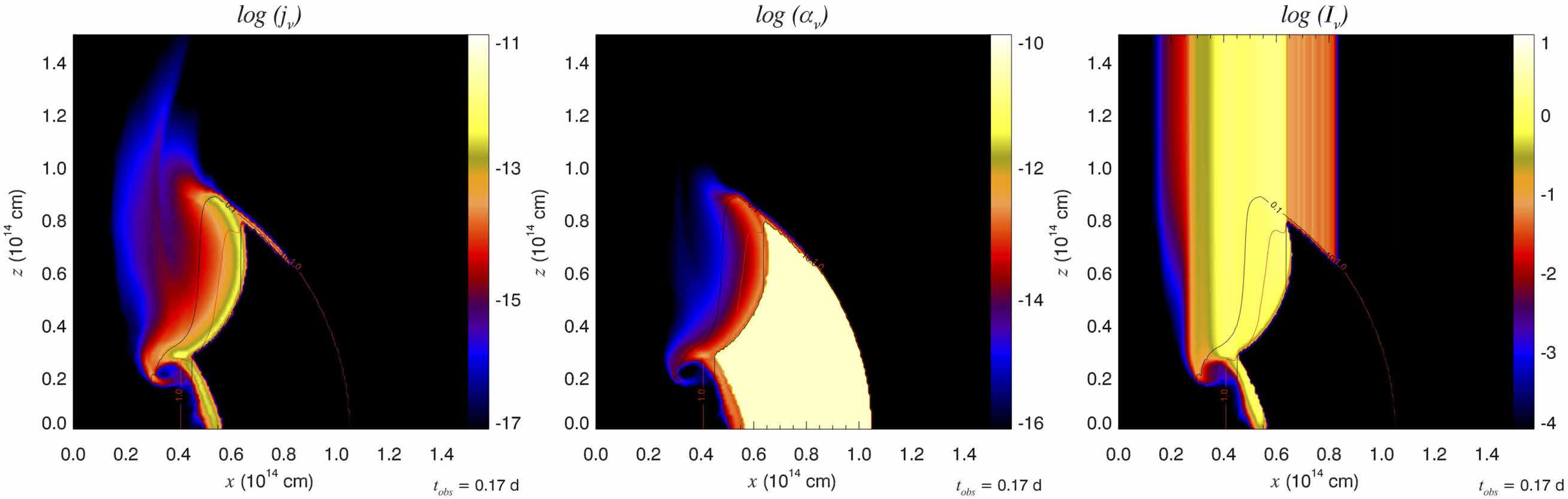}
\caption{Same as Fig.~\ref{fig:em-ab-in-obs} but for model G2.}
\label{fig:em-ab-in-obs-G2}
\end{figure*}
\begin{figure*}
\centering
\centering
\includegraphics[width=17.3cm]{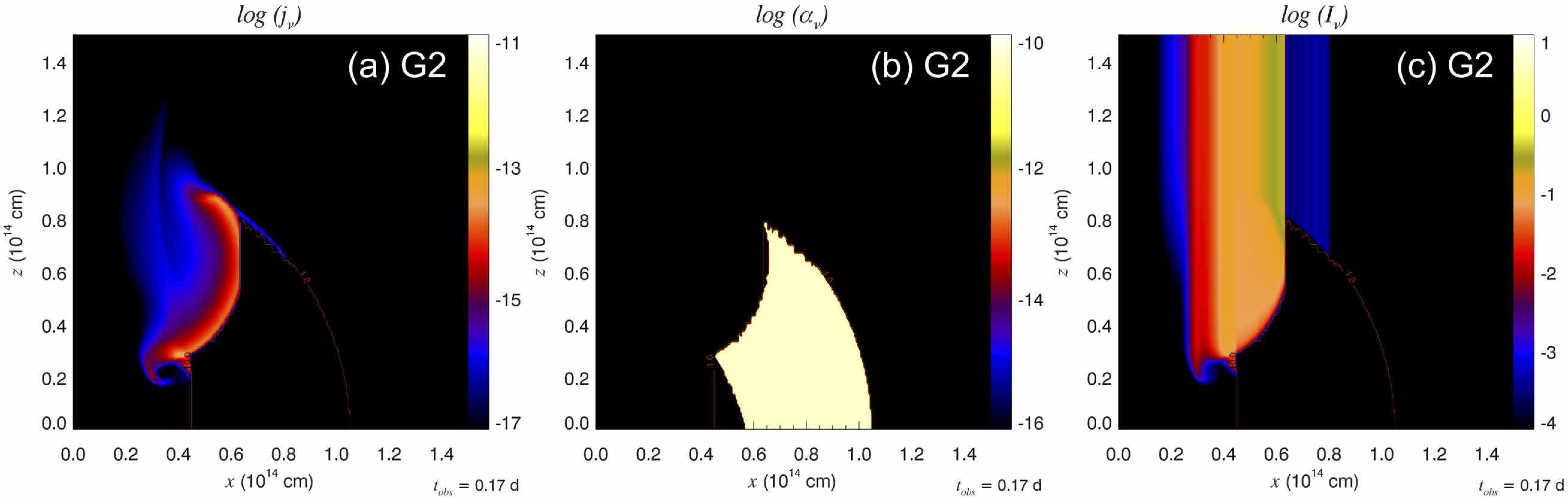}
\includegraphics[width=17.3cm]{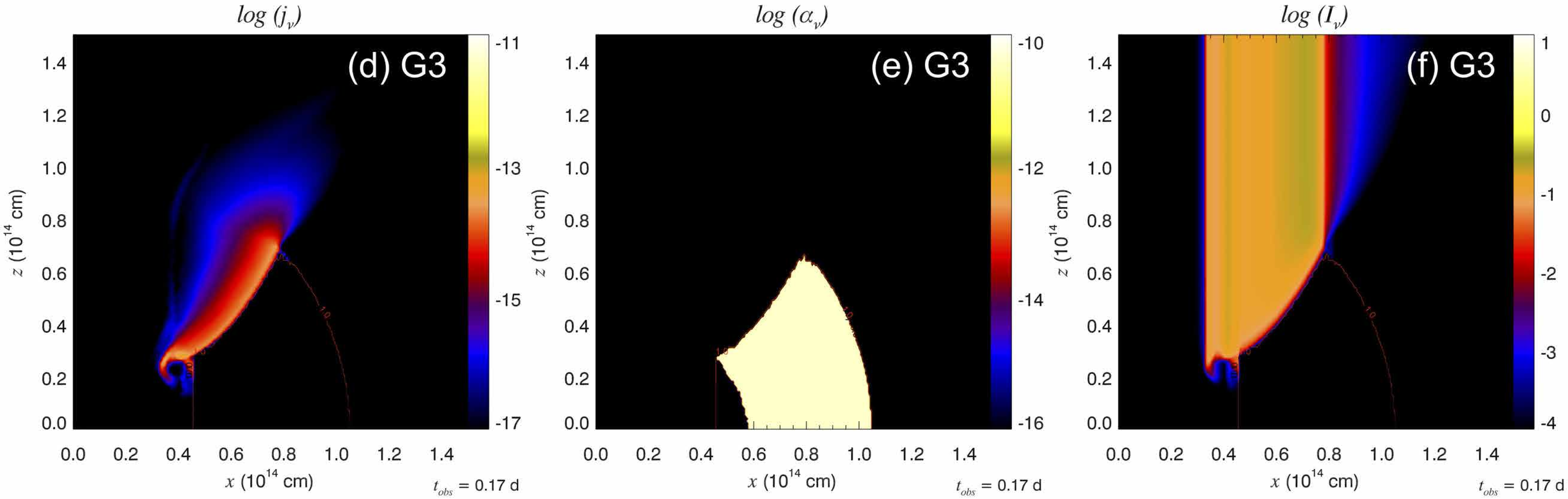}
\includegraphics[width=17.3cm]{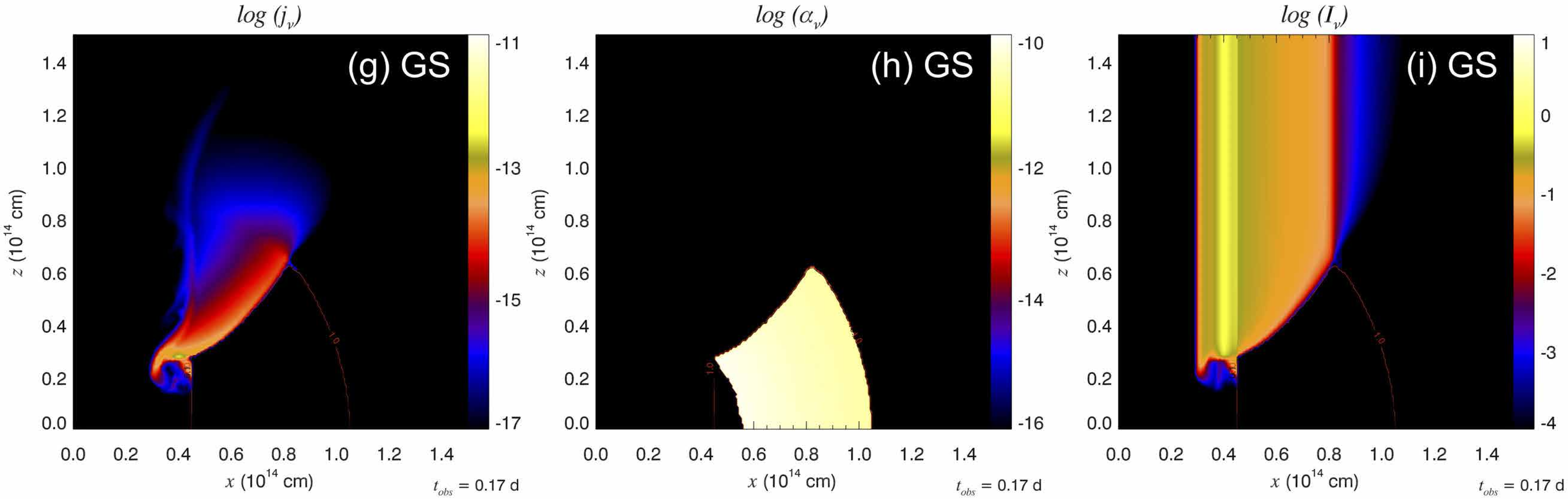}
\includegraphics[width=17.3cm]{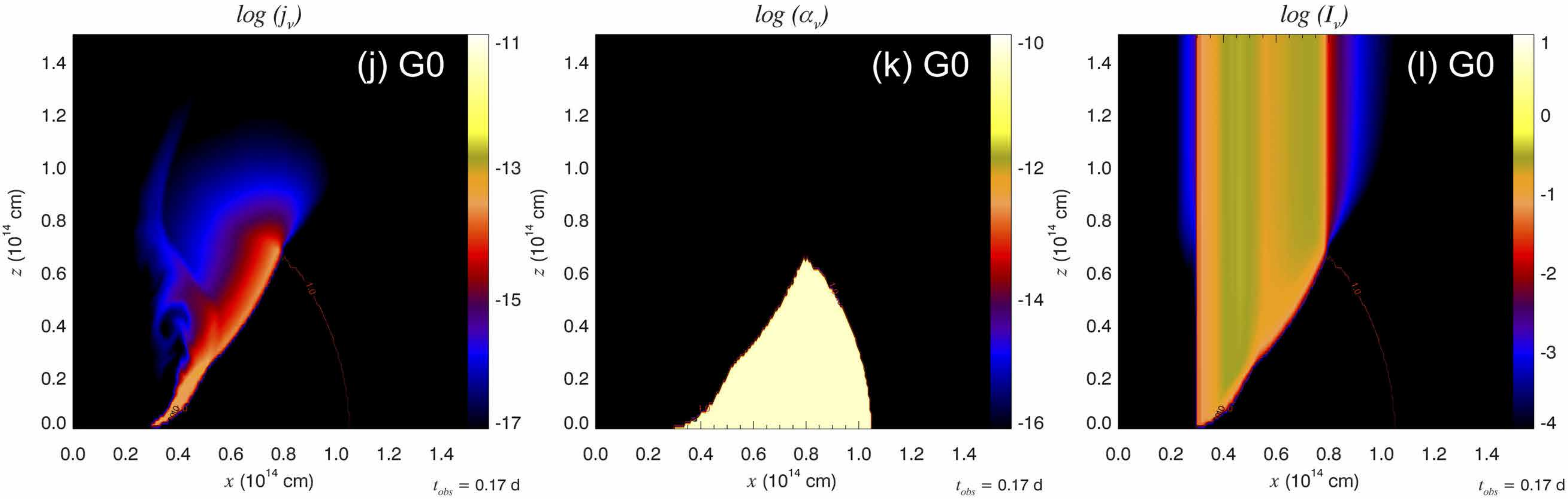}
\caption{Emission, $j_\nu$ (panels in the left-hand column) and absorption, $\alpha_\nu$ (central column) coefficients and evolution of the specific intensity, $I_\nu$ (panels in the right-hand column) along the line of sight. The observer is located in the vertical direction (towards the top of the page) at a viewing angle $\theta_{\rm obs} = 0^\circ$. The emission is computed in the X-ray band for free--free bremsstrahlung process, at an observational time of $0.17$\,d. Each of the rows corresponds to a different model:  G2 (top row), G3 (central upper row), GS (central bottom row) and G0 (bottom row).}
\label{fig:em-ab-in-obs-X-G2-G3-G1S}
\end{figure*}

Though most of the thermal radiation is emitted before $t_{\rm obs}\simeq 2\,$d from the jet/CE-shell interaction, there is also a minor thermal contribution originating from the expanding jet bubble that last much longer than the dominant thermal component. This contribution will depend on the properties of the EM as, e.g. its rest-mass density. Initially in the RM, the contribution to the observed flux of the bubble is two to four orders of magnitude smaller than the flux emerging from the jet/CE-shell interaction region. However, at later times ($t_{\rm obs} \gtrsim 5$\,d) the bubble emission still remains, and its thermal contribution tends to flatten the observed light curves. As we decrease the density in the EM we expect to have a less pronounced flattening, or even that the flattening does not show up during the time-scales of few days considered here. This is the case for model M2 where such flattening is absent after $\sim 3$\,d (Fig.~\ref{fig:reference-vs-stra}). For the stratified models S1 and S2, the flattening in the light curve is also absent. The reason is that the bubble density is smaller than in RM since the mass ploughed into the cavity at distances $R\sim10^{15}$~cm is smaller. Therefore, the bubble emission is also expected to be weaker. The emission for RM, M2, S1 and S2 models, in the three bands depicted in Fig.~\ref{fig:reference-vs-stra}, is practically the same until  $t_{\rm obs}\simeq 2\,$d, since the main thermal contribution is determined only by the CE-shell/jet interaction region.

The mass of the CE shell and the density contrast $\rho_{\rm CE,sh}/\rho_{\rm ext}$ play a key role shaping the emission properties of our models. As we have seen in Fig.~\ref{fig:referencethickthin} a CE-shell mass of $M_{\rm CE,sh} = 0.26 M_\odot$ suitably accommodates the observations (except at early times). We have tested other two models with different CE-shell masses. One with half the mass than the RM (model D2) and other with 10 times more mass (model D3). As we see in Fig.~\ref{fig:D2D3models} the latter model (dashed lines) leads to an emission peak which is two orders of magnitude above the observational data (or the upper limits) in the $W2$ band, and more than three orders of magnitude above the observations in the X-rays band. Furthermore, the emission peak at all frequencies is shifted to very late times, clearly incompatible with the observations. In model D3 the spectral reddening, if it happens, may take place after 5 d, i.e. too late to explain the observations. Contrarily, model D2 (solid lines in Fig.~\ref{fig:D2D3models}) shows peak fluxes (at all frequencies) at earlier times than in the RM, and the spectral inversion happens earlier than indicated by the observations. We therefore conclude that CE-shell masses not much larger than $\sim 0.26 M_\odot$ may account better for the observational data.

\subsection{CE-shell geometry}
\label{sec:CEgeometry}

By analysing the results of our simulations we found that most of the thermal contribution comes from the interaction region between the CE shell and the jet. Therefore, the exact details of the shell and funnel geometry can significantly influence the jet dynamics as well as the thermal emission. Thus, we have tested four different geometries of the CE-shell funnel to find out how they affect the emission: a toroidal geometry (RM, G3 and G0) and a simpler, linear geometry for the funnel (G2). A sketch of the two funnel geometries is displayed in Fig.~\ref{fig:shellgeometry}. The difference between models G3 and RM is that in the G3 model the funnel half-opening angle at $R_{\rm CE,out}$ is smaller than that of the RM. Model G0 differs from the RM in that it has a 10 times smaller rest-mass density in the EM, and the CE shell extends to the origin of the computational domain (i.e. there is no `gap' between the CE shell and the innermost radial boundary at $R_0$). We are using the same shell density in all these models. Therefore, the shell mass in the wedge spanned by the jet ($\theta_{\rm j}=17^\circ$ in the models considered here) is much smaller in the G2 model than in the case of a toroidal-like shell (RM). Likewise, the shell mass in the wedge spanned by the jet is larger in the G3 and in the G0 models than in the RM. Due to different funnel geometries, the CE-shell/jet interaction proceeds also differently. The presence of a `high'-density region close to the equator in model G2  (Fig.~\ref{fig:hydro-models}h) tells us that the CE-shell ablation process is not fully finished after $T\simeq 4\,$d in this model. This high-density region is not present in the toroidal cases  (Figs~\ref{fig:hydro-models}a and i). As a consequence of the smaller amount of swept up mass in the CE shell, the average bubble density (Fig.~\ref{fig:hydroDensity}) and mass (Fig.~\ref{fig:hydroMass}) are much smaller in the G2 model than in any of the other models presented here, at least, until $\sim 1\,$d. The late evolution of the G2 ($T \gtrsim 1\,$d) is akin to that of the RM, since the dynamics is then determined by the circumstellar medium. The break in the slope of the mass growth rate of model G2  (Fig.~\ref{fig:hydroMass}) is delayed with respect to most of the other models (it happens at $\sim 1\,$d). 

\begin{figure}
\centering
\centering
\includegraphics[width=8.4cm]{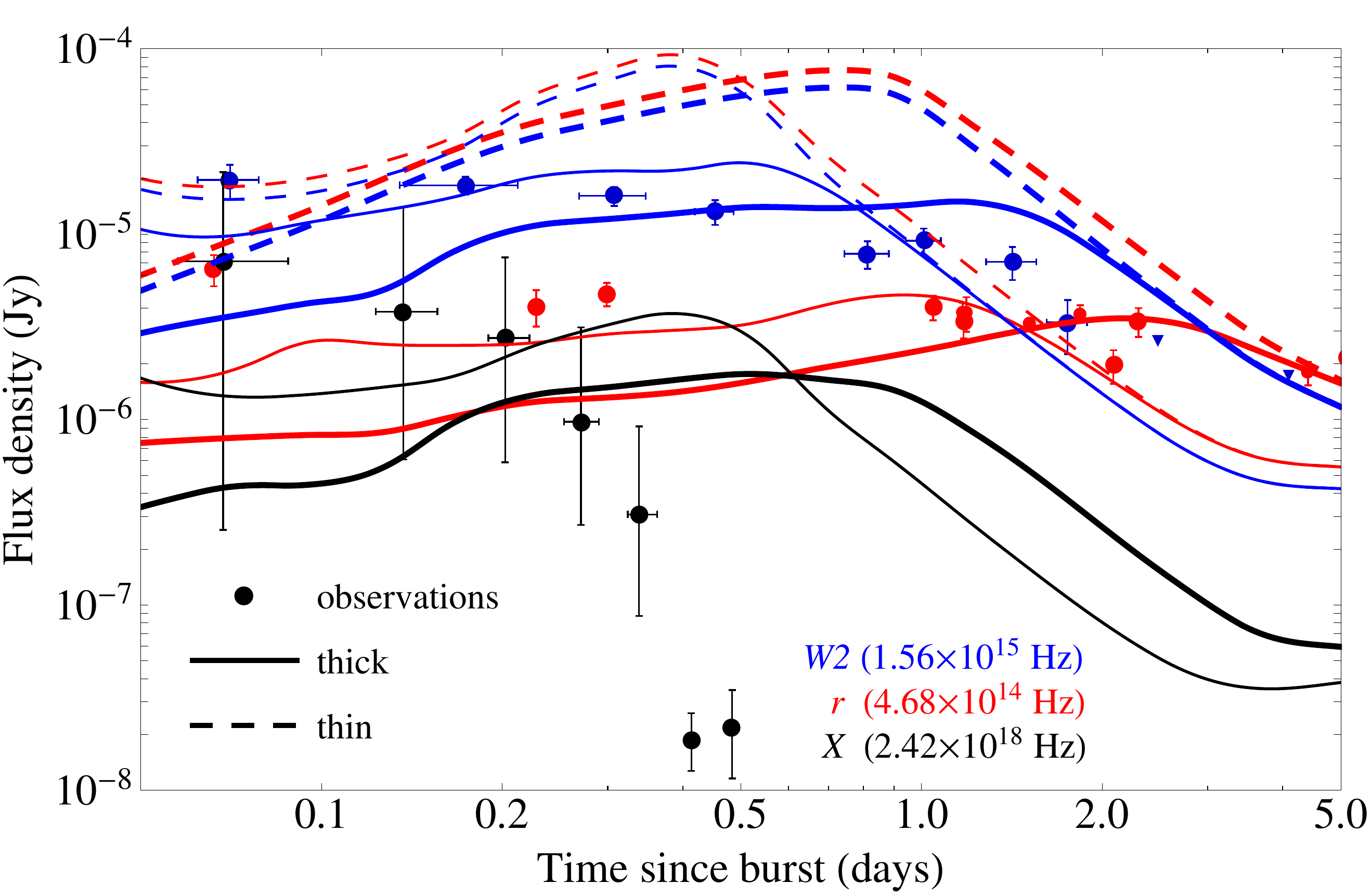}
\caption{Same as in Fig.~\ref{fig:referencethickthin} but for model G2 (thicker lines) and model G3 (thinner lines).}
\label{fig:G2thickthin}
\end{figure}

We have also computed the emission, absorption and specific intensity maps associated with the G2 model in the $W2$ (Fig.~\ref{fig:em-ab-in-obs-G2}) and in the X-ray (Fig.~\ref{fig:em-ab-in-obs-X-G2-G3-G1S} upper panels) bands. It is evident that, initially, the shape of the region from where most of the thermal emission is produced differs substantially between the RM and the G2 model. In the G2 model, the cross-sectional area of the thermally emitting region normal to the line of sight ($0^\circ$) is smaller (Fig.~\ref{fig:em-ab-in-obs-G2} left-hand panel and Fig.~\ref{fig:em-ab-in-obs-X-G2-G3-G1S}a) than in the RM. Such a shape is determined by the propagation of (forward and reverse) shocks sweeping the CE shell as the jet hits it. Because of the fact that, initially, the dominant emission region is much less inclined with respect to the line of sight in the case of model G2, the optical depth is also larger in such region, since radiation propagates upwards parallel to the symmetry axis and encounters denser parcels of the disrupted shell along the way. Indeed, we can observe the sharp cut-off in the specific intensity of the X-ray band of model G2 at about $5\times 10^{13}\,$cm from the symmetry axis and at $z\gtrsim 5\times 10^{13}\,$cm (Fig.~\ref{fig:em-ab-in-obs-X-G2-G3-G1S}c). This is associated with the very steep optical depth gradient in that region, as well as to a substantial decrease in the emissivity  (Fig.~\ref{fig:em-ab-in-obs-X-G2-G3-G1S}a), because there the fluid temperature is smaller.%
\footnote{This is a result of the model we have for the estimation of the fluid temperature from the total pressure including optical depth corrections (see Appendix~\ref{sec:temperature}).} Later in the evolution, the inclination of the emitting region with respect to the vertical direction grows, as the shocks resulting from the CE-shell/jet interaction sweep the CE shell towards the equator. This change in the inclination of the emitting region tends to reduce the optical thickness above it and to increase the effective emitting area, contributing, in part, to explain the delay in the peak flux at all frequencies when comparing the optically thick light curves resulting from thermal processes for models G2 (Fig.~\ref{fig:G2thickthin}) and RM (Fig.~\ref{fig:referencethickthin}). Also, until $\sim 0.7\,$d the flux in all the frequencies is smaller than in the RM, and falls below the observational data. Furthermore, there is an obvious deficit of thermal energy flux at early times in model G2. As in the RM, the system of model G2 is initially optically thick in the $W2$ and $r$ bands, but the transition to the optically thin regime happens later than in the former model (at $t_{\rm obs}\simeq 2\,$d in the $W2$ band and $t_{\rm obs}\simeq 4\,$d in the $r$ band).

All these features in the thermal emission result from the smaller amount of mass of the CE shell with which the jet is initially interacting, namely, the sector of the CE shell spanning from $\theta_{\rm f,in}$ to $\theta_{\rm j}$. For later reference, we will name this piece of the CE shell the CE-early-interaction wedge. Since the energy and momentum fluxes of the jet are the same in both models, the time needed to push away the CE-early-interaction wedge is smaller in model G2 than in the RM. Once the jet path is cleared, the jet/shell interaction weakens and, consequently, the time the jet needs to ablate the whole CE shell increases. This explains why the peak of the light curves at different frequencies is delayed in the G2 model with respect to the RM. It also explains why the initial thermal flux is smaller in the G2 model than in the RM, since the emitting region is also smaller in G2. Finally, the lower rate at which the CE shell is ablated in the G2 model leads to a delay in the transition to transparency in the $W2$ and $r$ bands with respect to the RM. 

Since the difference between the RM and model G3 is the cross-sectional radius of the central funnel (smaller in case of G3), the light curves of the RM (Fig.~\ref{fig:referencethickthin}) display smaller flux at early times ($t_{\rm obs} < 0.1$ d) and a peak at later times than those of the G3 model (Fig.~\ref{fig:G2thickthin}). In spite of this fact, light curves of the G3 and RM models are qualitatively more similar between them than those of model G2. The differences in the G3 model, with respect to the RM, arise as a result of the larger mass of the CE-early-interaction wedge in the former case. We also note that the pattern of the X-ray intensity distribution (Fig.~\ref{fig:em-ab-in-obs-X-G2-G3-G1S}f) in G3 is roughly similar to that of RM (Fig.~\ref{fig:em-ab-in-obs-X}).

From the comparison of the X-ray light curves in the RM  (Fig.~\ref{fig:referencethickthin}) and in the G2 and G3 models  (Fig.~\ref{fig:G2thickthin}), we note that our predicted flux in the X-ray band is very sensitive to the geometry and, more generally, to the physical conditions of the CE-early-interaction wedge. A higher CE-shell density close to the symmetry axis seems to {\em fit} the observational data better than a {\em wide} low-density funnel. We also note that the slope of the light curve after the X-ray maximum is very similar for all three models, and the same is true in the $W2$ band as well. 

\begin{figure}
\centering
\centering
\includegraphics[width=8.4cm]{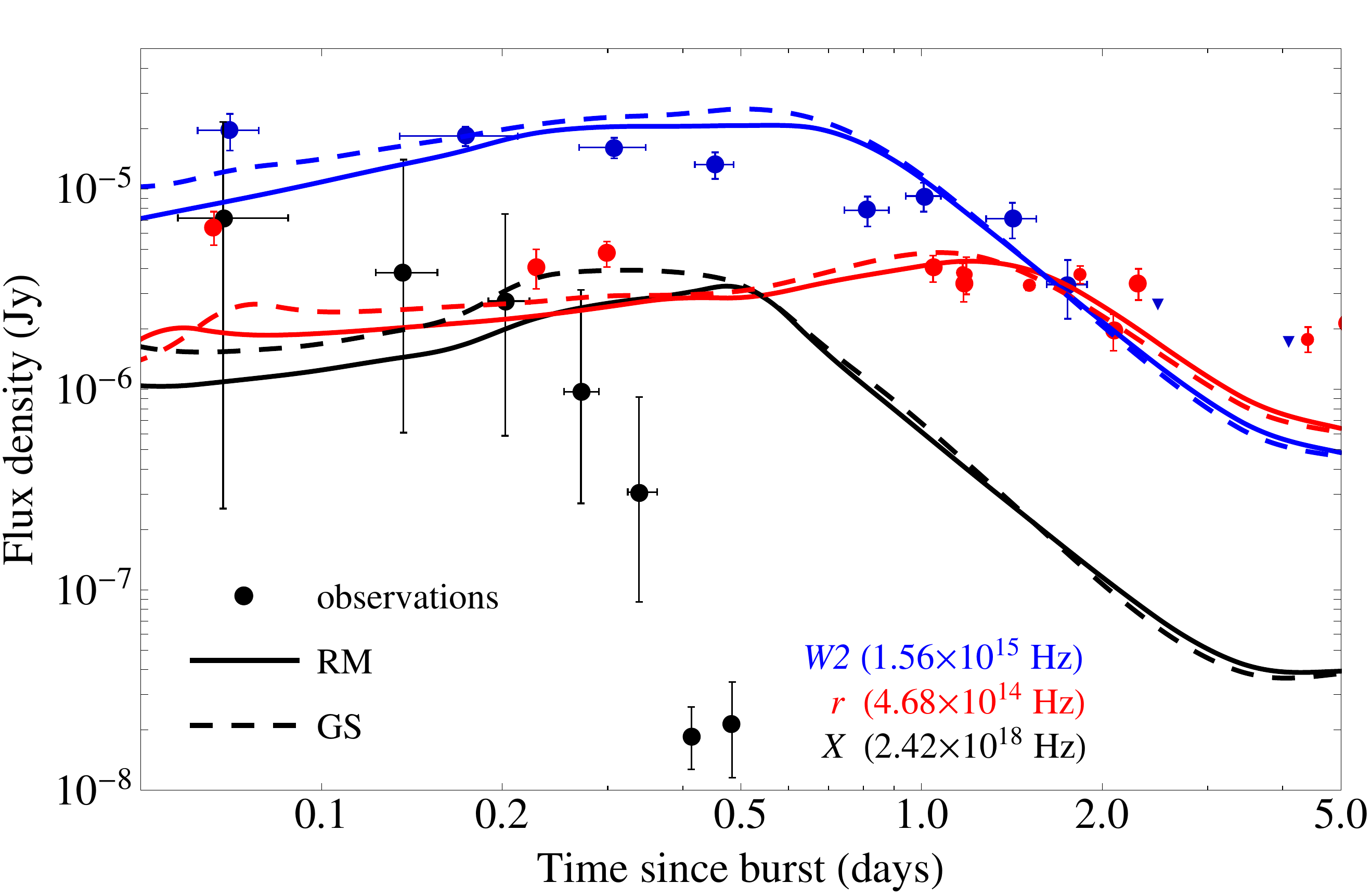}
\caption{Same as in Fig.~\ref{fig:referencethickthin} but for model GS (dashed lines) compared to the RM (solid lines). Only optically thick light curves are considered.}
\label{fig:GSthick}
\end{figure}

To better explain the observations a faster decrease after the maximum in the X-ray light curves is needed. This could be obtained by fine tuning the stratification of the CE shell. However, such a level of detail in the model set up is beyond the scope of this paper. Here we consider a simple stratification of the CE shell in which the rest-mass density and pressure decrease as $\propto r^{-2}$ (model GS). Comparing Fig.~\ref{fig:em-ab-in-obs-X} (right-hand panel; RM) and Fig.~\ref{fig:em-ab-in-obs-X-G2-G3-G1S}g (GS), we note that a stratified CE shell has a cross-sectional area of the X-ray emitting region which is similar to (though slightly larger than) that of the uniform CE shell of the RM. However, the specific intensity displays a stronger variation as we move away from the symmetry axis. In model GS, the specific intensity in X-rays is higher because of the contribution of (higher density) emitting regions which are closer to the symmetry axis. The change in the distribution of rest mass of the CE shell also yields small differences in the $W2$ and $r$-band light curves of model GS (Fig.~\ref{fig:GSthick}). We note that the observed flux in these two bands integrated up to the peak frequency in each band is $\lesssim 10$ per cent larger than in the RM, and that the peak at each frequency is shifted to a bit earlier times. After the maxima, the decay of the light curves is slightly faster than in the RM.

The differences between the RM and model G0 are very small, in spite of the fact that the CE shell in the later model extends down to the innermost radial computational domain, i.e. we take $R_{\rm CE,in} = R_0 = 3\times 10^{13}$\,cm. We can observe that the emission region in this case extends up to the innermost radial boundary of the computational domain (Fig.~\ref{fig:em-ab-in-obs-X-G2-G3-G1S}j). However, in spite of the small changes in the emission and absorption regions, the overall light curve of model G0 is almost indistinguishable from that of model M2 (Fig.~\ref{fig:reference-vs-stra}). G0 shows a slight increase of flux at early times, in the X-ray and $W2$ bands, since the jet/CE-shell interaction is a bit more extended. Provided that the only difference between the M2 and G0 models is the `gap' between the CE shell and the jet injection nozzle, we conclude that the effects of our specific initialization of the jet is negligible. Once again, this is a result of the fact that the thermal emission originates from the jet/CE-shell interaction and the small mass difference added in the CE shell of G0 with respect to models M2 or RM does not change neither qualitatively, nor quantitatively our results.

%
\subsection{X-ray emission}
\label{sec:X-rays}
\begin{figure}
\centering
\centering
\includegraphics[width=8.4cm]{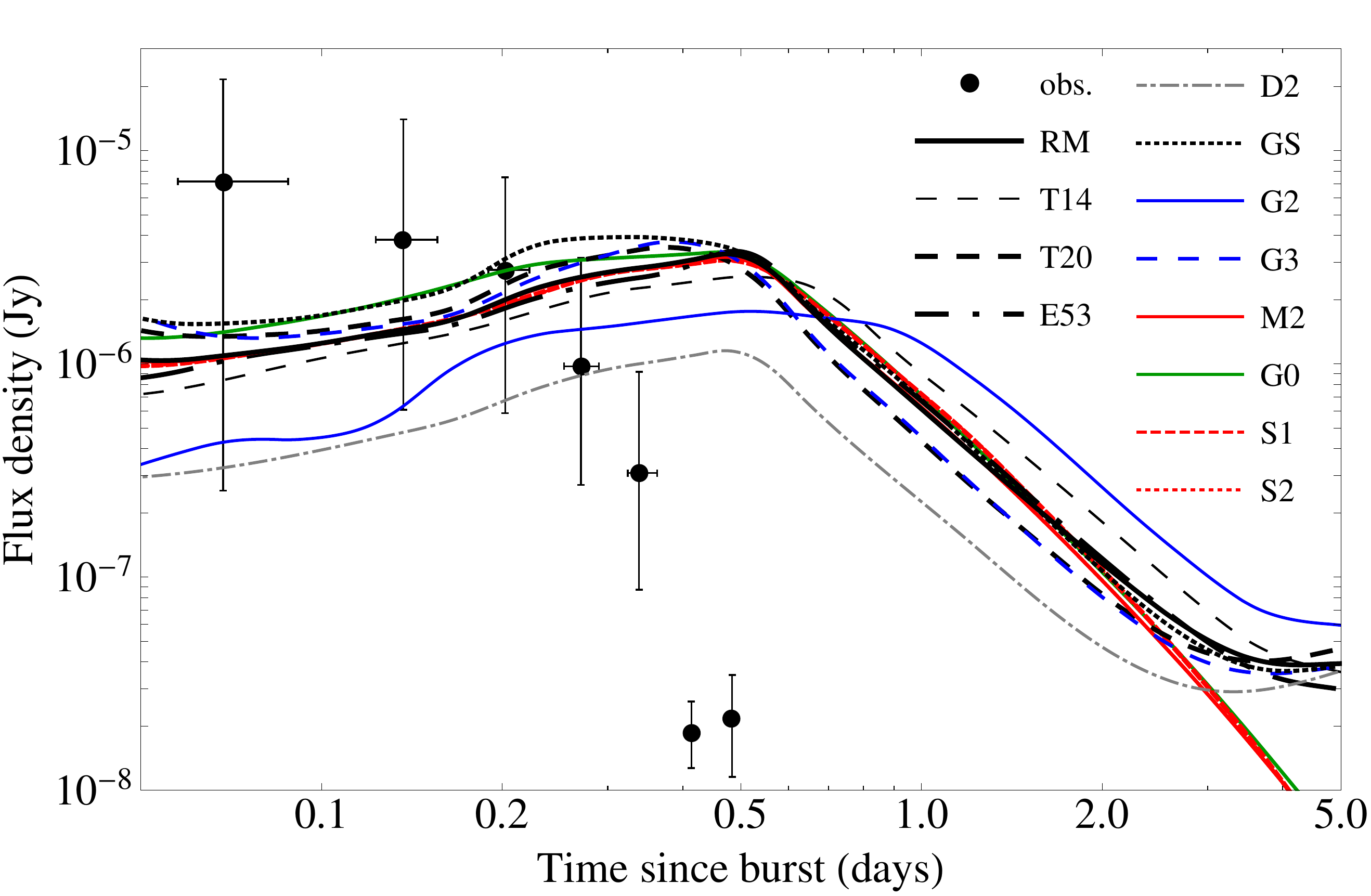}
\caption{Light curves for all the models in this paper (except D3) in the X-ray band. The X-ray data have been clustered as explained in Fig.~\ref{fig:referencethickthin}.}
\label{fig:X-rayLC}
\end{figure}

As we have seen in the previous section, the X-ray flux density in our models peaks too late with respect to the observations. In Fig.~\ref{fig:X-rayLC}, we display the X-ray light curves of all the models (except D3) in Table~\ref{tab:params}. The peak flux is model dependent: broader jets (T20) peak earlier ($t_{\rm X, peak}({\rm T20})\simeq 0.35\,$d) than the RM  ($t_{\rm X, peak}({\rm RM})\simeq 0.48\,$d) or narrower jets ($t_{\rm X, peak}({\rm T14})\simeq 0.6\,$d). The model which peaks latest is G2 (previously discussed  in Section~\ref{sec:CEgeometry}). The model D2 is the one with the lowest X-ray flux density. This is easy to understand since it is the model where the mass of the CE-early-interaction wedge is smaller, and where the CE-shell/jet interaction converts the smallest amount of kinetic into thermal energy. Changing only the stratification of the CE shell (compare models GS and RM in Fig.~\ref{fig:X-rayLC}) we realize that a stratified CE shell only increases the flux by factors of $\lesssim 2$ at early times, but after $\simeq 0.5\,$d the flux is very similar to that of the RM. G0 model shows a similar behaviour at early times since we decreased the innermost radius of the CE shell (keeping the opening angle of the funnel constant). Thus, as we have increased the CE-early-interaction wedge, the emission grows at early times. However, even more important than the small increase of the X-ray flux at early times is the fact that at late times, the X-ray light curve of model G0 (also of model M2; Fig.~\ref{fig:X-rayLC}) does not flatten, as a result of the 10 times smaller EM rest-mass density than in the RM.

In all the models, the flux density in X-rays until $\simeq 0.3\,$d is lower than the observations by approximately one order of magnitude. Integrating until $0.3\,$d the total flux density amounts to $\sim 10$--$30$ per cent of the observed flux. This result is broadly compatible with the analysis of T11, since they conclude that the X-ray hotspot displays a thermal component which accounts for $\sim 20$ per cent of the X-ray flux.  

We note that all of the models display an excess of X-ray flux density after $\sim 0.3\,$d. This can be improved by more sophisticated funnel geometries, since the geometry of the CE-shell funnel has an important influence on the X-ray peak time. However, we have not considered more complex funnel geometry to avoid increasing the number of free parameters in our models. Of course, the CE shells considered here are an oversimplified model of the very complex structures resulting from NS--He star mergers.

\section{Discussion and conclusions}
\label{sec:conclusions}

The CB has been interpreted by T11 as resulting from the merger of a NS with the He core of an evolved massive star. The key ingredient in that model is the ejection of the outer hydrogen layer of the secondary star, which adds a complex structure to the medium surrounding the progenitor system. In this paper, we have modelled the propagation of relativistic jets of different physical conditions through the outer layers of the secondary star and through the circumstellar medium, focusing on the jet/ejecta interaction dynamics. The ejecta are not the result of a self-consistent simulation of the merger of a NS with a He core. Instead, we parametrized the unbound CE matter as a shell that, by the time the ultrareltivistic jet catches up, has expanded out to $\gtrsim 10^{14}\,$cm. To assess the reliability of our results we have performed a parametric scan of the most important physical properties of the jet (by varying $E_{\rm iso}$ and $\theta_{\rm j}$), of the CE shell (by varying its rest-mass density and its geometry), and of the circumburst medium (by considering either uniform or stratified cases). The parametric scan has been performed via several numerical two-dimensional, axisymmetric, special relativistic simulations. The simulations support the idea presented in T11, and explain the bizarre phenomenology of GRB 101225A, in particular and, by extension, of the so-called BBD-GRBs. Using a full radiative transport code, \tiny{SPEV}\normalsize, we post-process the previously computed hydrodynamical models run with the relativistic (magneto)-hydrodynamics code \tiny{MRGENESIS }\normalsize and estimate their synthetic thermal (free--free bremsstrahlung) emission signature. The numerically computed emission is compared with the first 5 d of UVOIR and X-Ray Telescope (XRT) observations.

All simulated jets and ejecta undergo a very similar dynamical evolution that can be divided up in three stages. In the first phase, an ultrarelativistic jet is injected through a small nozzle at a distance $R_0=3\times 10^{13}\,$cm and freely expands until hitting the inner surface of the CE-ejecta shell.\footnote{This first stage is absent in the model G0, where we do not set up any gap between the CE shell and the innermost radial boundary of our domain, in which case, the dynamics begins directly in the second phase we describe in the text.} The second phase begins when the jet encounters the funnel in the ejecta. Since the ejecta have a toroidal structure, with a funnel along the symmetry axis, and since the jet is broader than the ejecta funnel, a minor fraction of the jet  (its central core) proceeds through the funnel. Simultaneously, the outer layers of the jet impact against the CE shell, much denser than the EM. Because the geometry of the CE shell is non-trivial, a number of oblique shocks result from the CE-shell/jet interaction. Simplifying the picture, we may say that two types of shocks form  as a result of the interaction. They propagate at a certain angle with respect to the radial direction (i.e. with respect to the direction of propagation of the jet). Some of the shocks sweep the CE shell and heat it up, while other shocks move towards the jet axis and convert a fraction of the jet kinetic energy into thermal energy. Furthermore, the jet progressively displaces and pushes forward the fraction of the CE shell which is on its path. This is the mechanism by which the jet accumulates substantial baryonic mass, so that a very quick deceleration process begins. The jet injection lasts for $\sim 3800\,$s, a time by which the head of the jet core breaks out of the outer boundary of the CE shell (located at $R_{\rm CE,out}\simeq 10^{14}\,$cm). Although, for numerical reasons, the jet injection does not immediately ceases at $\sim 3800$\,s, after that time the amount of energy still injected is tiny (the jet luminosity decreases as $t^{-4}$). 

In the third stage, the baryon-loaded, shock-heated jet inflates a cavity, which is initially prolate.  The evolution after the first day enters into a quasi-self-similar regime, so that the aspect ratio (i.e. the cross-sectional to longitudinal bubble diameter ratio) grows monotonically. Extrapolating our results indicates that the shape of the cavity, propagating in a uniform, high-density EM, will be roughly spherical after $12$--$18\,$d. At this stage, the cavity expansion rate is monotonically decreasing. The speed of propagation of the outer edge of the bubble is mildly relativistic and decreases from $\simeq 0.9c$ (after $\simeq 0.1\,$d) to subrelativistic values $\simeq 0.1c$ (after $\simeq 4\,$d). Starting from the end of the second phase (roughly coincident with the time at which we stop the jet injection) the initial jet structure is progressively being disrupted and, after a few hours a jet beam cannot be identified any more.

The cavity dynamics in the third evolutionary stage strongly depends on the EM characteristics. It is chiefly determined by the balance between the EM mass ploughed by the outermost (forward) shock, and the energy injected into the cavity by the jet. Within the first 4 d of evolution our models accumulate $\sim 1$--$2 M_\odot$ from the EM (when it is assumed to be uniform and with a high density). Since the external shock is subrelativistic, it does not leave the typical fingerprint of a standard GRB afterglow. However, as we show in the companion Paper II, this shock has a non-negligible signature in the UVOIR bands. During the third evolutionary stage, the CE shell is fully ablated by the shocks triggered during the second stage of evolution. These shocks transfer momentum to the CE shell and heat it up. After $\sim 2\,$d, the whole CE shell is disrupted and has expanded significantly, lowering by a factor of $10$--$100$ the rest-mass density of the region initially occupied by the CE shell.

On top of the basic evolutionary dynamics described above, we find a number of differences between models, specially during the first hours of evolution. The CE-shell/jet interaction is a non-linear process which depends on the jet energy and its angular extent. Broader jets increase the effective interaction region and incorporate more mass from the CE shell than narrow ones. More energetic jets blow the jet cavity faster, and ablate the CE shell earlier. Additionally, the jets propagating into lower density EM develop more prolate cavities. These tend to adopt a spherical shape later than those of the higher density external media. Another key factor shaping the CE-shell/jet interaction is the CE-shell funnel geometry. As stated above, our model of the CE ejecta stems from the results of past simulations. Here we have considered a simplified (linear) funnel structure and a (more elaborate) `toroidal' funnel geometry, where the funnel half-opening angle grows non-linearly from a minimum value at the radial inner face of the CE shell ($\theta_{\rm f,in}$) to a maximum one ($\theta_{\rm f,out}$; see Fig.~\ref{fig:shellgeometry}). These two funnel geometries change substantially the amount of mass of the CE shell which is within reach of the relativistic jet (the angular region $[\theta_{\rm f,in}, \theta_{\rm j}]$, and the radial region $R_{\rm CE,in}$ to $R_{\rm CE,out}$. This region is very quickly incorporated into the jet beam and contributes to the early jet deceleration. Those shell geometries in which there is a large amount of rest mass close to the symmetry axis maximize the CE-shell/jet interaction and decelerate the jet beam more rapidly. Finally, reducing the rest-mass density of the CE shell the jet is decelerated more slowly, the CE shell is ablated sooner, and the average cavity rest-mass density is smaller. However, these effects do not translate into a very different bubble evolution because most of the cavity mass does not come from the CE shell, but from the EM, which is the same in most of the models we considered.

Apart from the jet/shell interaction dynamics, the landmark of this paper is the identification of the CE-shell/jet interaction region as the origin of the thermal emission. We find that the UVOIR observations can be chiefly explained as radiation coming from the CE-shell/jet interaction region, rather than the surface of an expanding bubble as proposed in T11. The overall contribution of the expanding hot bubble to the total observed thermal flux is negligible during the first 4 d of evolution. This is specially true in the models with the lower rest-mass density in the EM. According to our models, the region from where most of the thermal emission comes from is much smaller ($\lesssim 5\times 10^{13}$ cm) than the size of the cavity blown by the simulated jets (with a size that grows up to $\lesssim 10^{15}\,$cm). Furthermore, the thermal emitting region in the UVOIR bands is a mixture of transparent and semitransparent regions. These conclusions do not change if we consider a different EM, as a more realistic stratification of the rest-mass density and of the pressure. Though the dynamical differences are apparent, the thermal emission does not differ much with respect to the corresponding models with a uniform EM. The reason is that, as stated above, the origin of the thermal emission is the jet/shell interaction, and, models having the same CE-shell and jet parameters yield very similar thermal light curves.

The agreement with observational data in the UVOIR bands is not optimal during the first $\simeq 0.2\,$d, in which we underpredict the observed flux by a factor of $\lesssim 3$. However, it needs to be considered that observations also include a non-thermal contribution, which may account for the flux deficit at early times (see Paper II).

The spectral reddening, which produces a spectral inversion in optical bands between $1.5$--$2$ d, is caused by the transition from an optically thick to an optically thin emitting regime. The dynamical reason for such a transition is the complete ablation of the CE shell by the outward pushing ultrarelativistic jet.

As anticipated by T11, there are three key elements that any theoretical model of the progenitor of the CB must explain: the persistent X-ray hotspot, the lack of a standard afterglow, and the UVOIR BB evolution. The current special RHD models provide a likely explanation, for two of these features, namely, the origin of the thermal emission in the subclass of BBD-GRBs and the effective absence of a classical afterglow. The thermal signal results from the interaction between the jet and the CE shell ejected in the late stages of the progenitor system evolution. The suppression of a classical afterglow happens because the relativistic jet is baryon polluted as it interacts with the CE shell. 

In the X-ray band,  T11 conclude that $\sim 20$ per cent of the flux can be attributed to a bright hotspot, radiating as a BB at a temperature $\simeq 1$--$1.5\,$keV until $\lesssim 0.34\,$d. The X-ray flux estimated from our models is marginally consistent with such an observational fit until $\lesssim 0.3\,$d. Furthermore, during that period of time the brightness temperature we infer from our models of the X-ray emission is $\simeq 0.6$--$0.7\,$keV.  However, we overpredict the duration of the X-ray emission, whose maximum happens in our models $\sim 0.5$--$0.7\,$d after the GRB. After $\simeq 0.3\,$d, our models overestimate X-ray flux. The main reason for the discrepancy is that the X-ray emission comes from a significant fraction of the CE-shell/jet interaction region. This region is much larger ($\lesssim 4\times 10^{13}\,$cm) than the size estimated for the X-ray hotspot in T11 ($\simeq 2\times 10^{11}\,$cm). However, our simulations indicate that radiative flux has a very strong dependence on the geometry of the CE-shell funnel. The `toroidally' shaped funnel seems to reduce X-ray flux. Therefore, extrapolating our results, we suggest that an initially narrower funnel with a larger density would improve the results obtained in the X-ray band. Unfortunately, this would require substantially increasing the numerical resolution of our models close to the symmetry axis, and likely, extending the CE shell towards smaller radii. Both facts drastically reduce the time step with which we shall run our models and notably enlarge the execution time. We defer this study to a future work.

As a cautionary note, we want to outline that our models support the possibility that BBD-GRBs are the byproduct of NS/He mergers, but do not rule out other  possible progenitor models for BBD-GRBs. Our models assume a specific distribution of rest mass in the EM surrounding the progenitor. NS/He mergers provide such structure naturally, but we may not discard the possibility that in single-star progenitor models (e.g. \citealt{Nakauchi_etal_2013ApJ...778...67}) the original massive star ejects mass non-isotropically. If this mass is preferentially ejected along the equatorial regions and leaves a relatively low-density funnel around the rotational axis of the system, we foresee that the jet/EM interaction would result in a dynamics qualitatively similar to the one described in this work. Hence, we also preview a non-trivial thermal signature in the latter case.

Our models accommodate better the observational data if the GRB-jet (true) energy is $\lesssim 10^{52}\,$erg. According to \cite{Fryer_etal_2013ApJ...764..181}, this places some (soft) restrictions of the mass of the He cores that shall merge. For neutrino-powered jets, it would be requested that the He core mass be larger than $\sim 10 M_\odot$. However,  if the jet is magnetically powered, the previous restriction is not so stringent in practice, since for He cores with masses $\gtrsim 3 M_\odot$ one may get a sufficient amount of energy.

The fact that our models with a large CE-shell mass ($M_{\rm CE,sh}\simeq 2.6 M_\odot$) is at odds with observations is suggestive of several possibilities: (1) that the secondary star of the merger has either a relatively small He-core mass, (2) that most of the envelope of the secondary has been ejected before the final CE phase begins, or (3) that the fraction of the CE participating in the thermal emission during the first 2 d is relatively minor compared with the rest of the (likely bound) CE interior to the CE shell. Elucidating which of these possibilities is more likely is not possible with our models, since they depend on the initial mass distribution in the envelope of the star, i.e. on the details of a region not included in our models.

Finally, we point out that in this paper, we have focused on the dynamics and the origin of the thermal emission in BBD-GRBs. However, our jet models develop shocks, specially the bow (forward) shock surrounding the blown up cavity, where non-thermal (synchrotron) emission shall be produced. In the companion Paper II we address such non-thermal contribution to the total emission of these events. We anticipate that this contribution is significant at early times, when the jet has not been fully disrupted yet.

\section*{Acknowledgements}
We acknowledge the insightful discussions and useful comments of A. de Ugarte-Postigo and C. Th\"one. 
We acknowledge the support from the European Research Council (Starting Independent Researcher Grant CAMAP-259276), and the partial support of grants AYA2010-21097-C03-01, CSD2007-00050 and PROMETEO-2009-103. CC-M also acknowledges the support of ACIF/2013/278 fellowship, and the partial support of UV-INV-PREDOC13-110509 fellowship. We thankfully acknowledge the computer resources  (`Llu\'is Vives' and `Tirant' supercomputers), technical expertise and assistance provided by the Servei de Informatica at the University of Valencia.

\appendix

\section{Thermal emission}
\label{sec:spev}

We have implemented a new algorithm for computing (thermal) bremsstrahlung-BB radiation with \tiny{SPEV}\normalsize. In the following we provide some of the details about the method, for which we first need to evaluate emission and absorption coefficients at each numerical cell (Appendix~\ref{sec:emis+abs}) and, since we are using a simplified equation of state (the \emph{TM} approximation), we also need to provide a method to compute the temperature of each cell considering its local thermodynamic properties (pressure and rest-mass density), as well as the optical depth (Appendix~\ref{sec:temperature}).

\subsection{Emission and absorption coefficients}
\label{sec:emis+abs}

The radiation transport equation shows how the intensity per unit frequency, $I_\nu$, changes because of emission and absorption processes specified through the coefficients $j_\nu$ and $\alpha_\nu$, respectively, along the path, $s$, of the photon: 
\begin{equation}
\frac{dI_\nu}{ds} = - \alpha_\nu I_\nu + j_\nu.
\label{eq:radiativetransport}
\end{equation}
Here photon paths are straight lines, since we neglect GR effects.

Through the computational domain we assume that only fluid elements above a certain threshold in temperature, $T_{\rm  th} = 25000$\,K, or above a certain threshold in velocity, $v_{\rm th} \simeq 0.00045 c$, emit thermal radiation. At least one of these conditions is fulfilled in every fluid element inside the relativistic jet, and inside the interaction region between the jet and the CE shell. However, imposing the former thresholds, we avoid computing the emission from the {\em cold} CE shell itself, where $T_{\rm CE,sh} \sim 19900$\,K $< T_{\rm th}$. The velocity threshold, avoids including the absorption of the EM, which we have ignored here for simplicity. The CE shell is, however very important because of its absorption properties. In the density/temperature conditions of the CE shell, it acts as an Thompson absorber, having a grey absorption coefficient $\alpha_{\rm t} = 0.2 (1 + X) \rho$ cm$^{-1}$.

In the following we describe the emission and absorption coefficients \citep{Rybicki_Lightman_1979rpa..book.....} of free--free thermal bremsstrahlung. First of all, we define the dimensionless variable 
\begin{equation}
x = \frac{h\nu}{k T},
\label{eq:x}
\end{equation}
where $h$ is the \emph{Planck} constant, $\nu$ is the frequency of the radiation, $k$ is the Boltzmann constant and $T$ is the temperature of the fluid in the comoving frame. For a plasma with a Maxwellian distribution of velocities the emission coefficient per unit of frequency, $\nu$, takes the form
\begin{equation}
j_\nu =  5.4 \times 10^{-37} Z^2 \frac{\rho^2}{m_{\rm p}^2} T^{-1/2} e^{-x} \bar{g}_{\text{ff}} (\nu,T)\, \text{erg s$^{-1}$ cm$^{-3}$ Hz$^{-1}$}. 
\label{eq:em}
\end{equation} 

Following the Kramer's law for opacity, the absorption coefficient per unit frequency is determined by the relation $\alpha_\nu = j_\nu / B_\nu$, where $B_\nu$ is the BB intensity. Then we have
\begin{equation}
\alpha_\nu \simeq 4.1 \times 10^{-23} Z^2 \frac{\rho^2}{m_{\rm p}^2} T^{-7/2} x^{-3} (1-e^{-x}) \bar{g}_{\text{ff}} (\nu,T)\, \text{cm$^{-1}$}.
\label{eq:abs}
\end{equation} 
In the previous two expressions, $\rho$ is the rest-mass density, $m_p$ the proton mass, $\bar{g}_{\text{ff}}$ the Maxwellian averaged Gaunt factor for free--free transitions and $Z=\mu_{\rm i}/\mu_{\rm e}$, where $\mu_{\rm e} = 2 / (1 + X)$ and $\mu_{\rm i} = 4 / (1 + 3 X)$. The variable $X$ is the relative abundance of hydrogen and we have chosen a typical value $X = 0.71$.

The Maxwellian averaged free--free Gaunt factor has been obtained by interpolation of the values computed by \citeauthor{Sutherland_1998MNRAS.300..321} (1998; see table 2 within), that depend on the variables $x$ (defined as $u$ in the paper) and $\gamma^2 = Z^2\text{Ry}/kT$, where Ry is the Rydberg energy.
In our range of temperatures and frequencies the Gaunt factor is close to unity, in which case $\bar{g}_{\text{ff}} (\nu,T) \sim 1$. As an alternative to the table interpolation chosen here, we could have approximated by 1 the Gaunt factor at low temperatures and take the expression proposed by \cite{Anderson_etal_2010PhRvD..81d4004} for high temperatures ($T > 10^5$\,K). It should be remarked that \citeauthor{Anderson_etal_2010PhRvD..81d4004}'s expression 9 contains a typo since $x$ is used as $k T / h \nu$ instead of its initial definition given by their equation~6, in which $x = h \nu / k T$ (compare with our equation~\ref{eq:x}). Interested readers can check equation~8 in \cite{Shapiro_Knight_1978ApJ...224.1028}, for comparison with equation~9 in \cite{Anderson_etal_2010PhRvD..81d4004}.

\subsection{Temperature calculation}
\label{sec:temperature}

The temperature of the electrons present in the fluid has been computed assuming that matter is coupled with radiation when the optical depth, $\tau$, is large enough. In that case the equilibrium temperature can be obtained numerically  (using Newton--Raphson method) from the following equation for the total pressure, which takes into account contributions of both electron and radiation pressure,
\begin{equation}
P =  P_{\rm e,bar} + P_{\rm rad} = \frac{k}{\mu m_{\rm H}} \rho T + \frac{1}{3} a T^4\, .
\label{eq:temperature}
\end{equation}
Here $m_{\rm H}$ ($\approx m_{\rm p}$) is the mass of the hydrogen atom, $\mu = (1 / \mu_{\rm e} + 1/\mu_{\rm i})^{-1} \approx 4 / (3 + 5X)$ is the mean molecular weight in units of $m_{\rm H}$, and $a$ is the radiation constant.
 
In the parcels of fluid which are not optically thick, i.e. where the optical depth is small, we assume that radiation is partially decoupled from matter. Thus, in a simple generalization of equation~(\ref{eq:temperature}), the total pressure is computed as
\begin{equation}
P =  P_{\rm e,bar} + P_{\rm rad}  (1 - e^{-\tau_{\rm t}}).
\label{eq:total-pres}
\end{equation}
Here $\tau_{\rm t}$ is the total optical depth of the system, along the line of sight, computed from the optical depth in each cell of the simulation. The latter is given by $\tau = \tau(T, \nu) =  \alpha_\nu (T, \nu) l$, with $l$ being the length of a cell along the line of sight.

As $\tau$ depends on the temperature, and for computing properly the temperature we must know the overall optical depth of our model, we need to perform an iterative process, which is repeated until a desired convergence is reached. For the initial guess of the temperature we assume that matter and radiation are coupled. We remark that the temperature depends on frequency, so we have different temperatures for different frequency bands.

We note that using this method for computing the temperature, in regions where there is a large gradient in optical depth, we may find also a very large temperature gradient, so that optically thick regions are much cooler than optically thin ones.

\section{Resolution study}
\label{sec:resolution}
\begin{figure*}
\centering
\includegraphics[width=17.6cm]{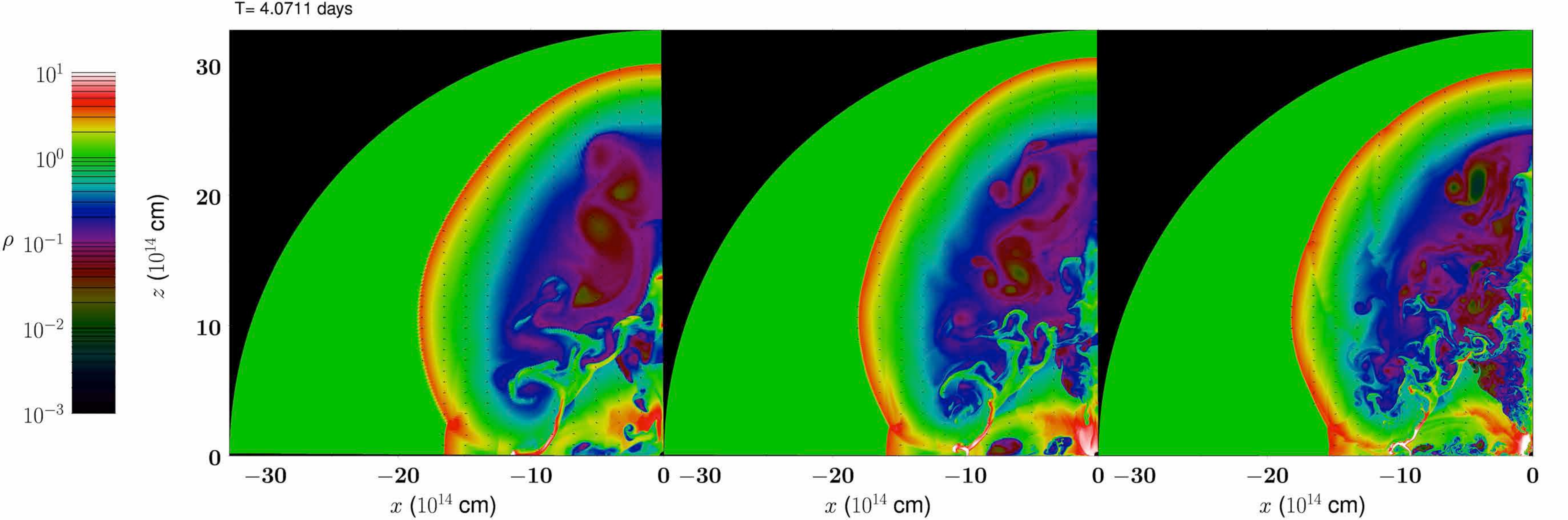}
\caption{Snapshots, at the end of the simulation, of the same model with three different mesh sizes, $n_{\rm r} \times n_\theta$ = $2700\times135$ (left), $5400\times270$ (centre) and $10800\times540$ (right).}
\label{fig:resolution}
\end{figure*}

We have performed a resolution study in order to test the convergence of the morphological evolution of the outflows and select an adequate mesh spacing. We have tested three different sizes: $n_{\rm r} \times n_\theta$ = $2700\times135$ (low resolution), $5400\times270$ (standard resolution; employed in all the models listed in Table~\ref{tab:params} which have a uniform medium) and  $10800\times540$.

\begin{figure}
\centering
\centering
\includegraphics[width=8.4cm]{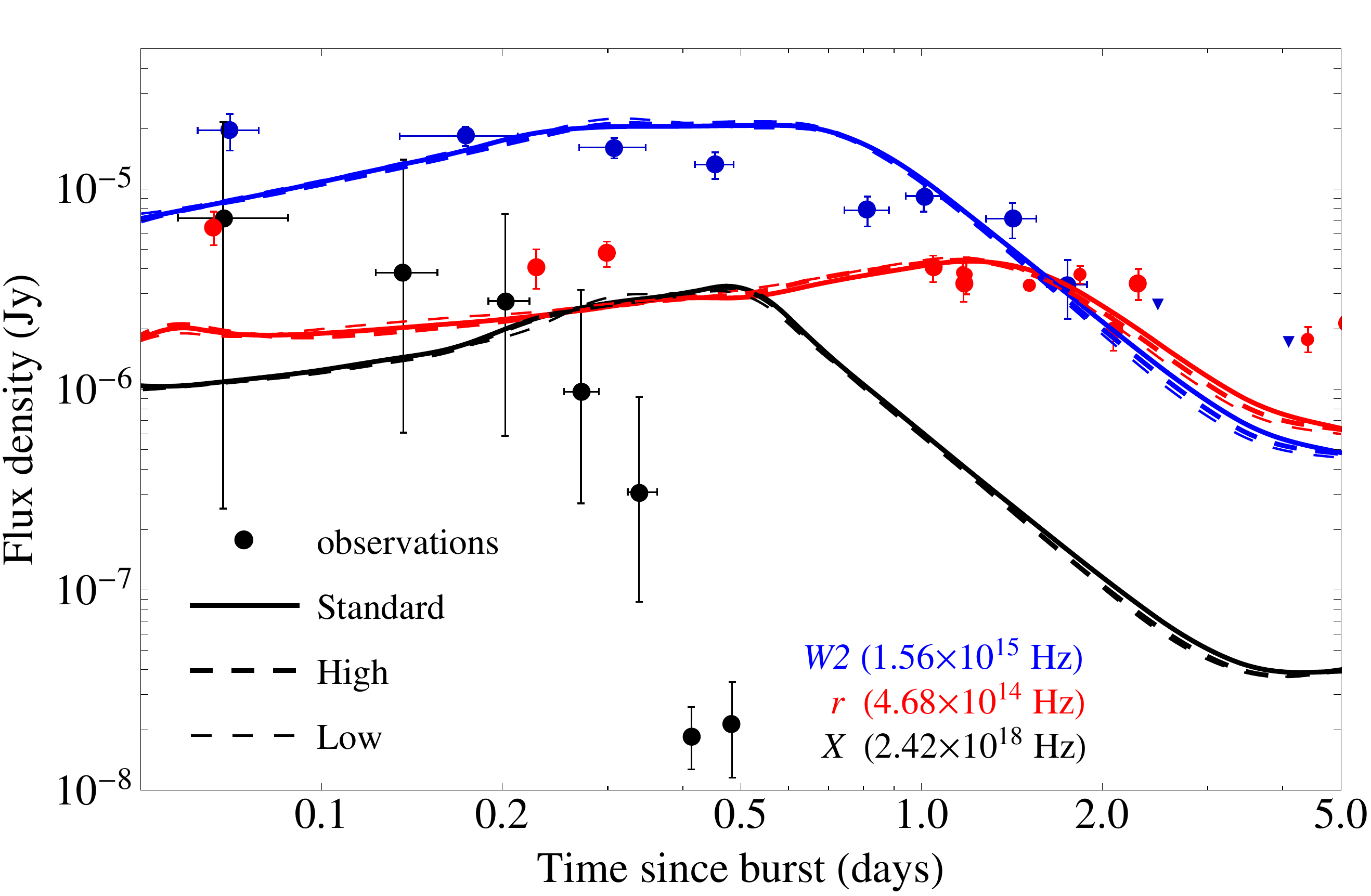}
\caption{Light curves for the RM considering only the (thermal) bremsstrahlung-BB contribution, comparing the convergence of the results employing different mesh sizes, $n_{\rm r} \times n_\theta$: $5400\times270$ (standard; solid lines), $10800\times540$ (high; thick dashed lines) and $2700\times135$ (low; thin dashed lines).}
\label{fig:resol-LCs}
\end{figure}

As we can see in Fig. \ref{fig:resolution} the morphological evolution of all three cases is reasonably similar; the jet head has reached the same position in the $z$-axis in all the cases. The transverse expansion of the outflow is also consistent. Obviously, the exact morphology of the turbulent internal part of the cavity is not the same, but the exact details of such region are irrelevant to shape either the non-thermal emission, dominated by the bow and reverse shocks, the properties of which are very similar in the standard and high-resolution runs, or the thermal emission, dominated by the jet/CE-shell interaction. To show that the thermal emission is roughly the same in all three cases we have explicitly computed the light curves due to thermal emission processes for the different resolutions (Fig.~\ref{fig:resol-LCs}). The fact that the synthetic emission depends only weakly on the resolution is because the jet/CE-shell interaction region is sufficiently well resolved in all cases. Therefore, we are justified in choosing a mesh size $n_{\rm r} \times n_\theta = 5400\times270$ (standard resolution) for all our simulations, because it gives the best trade-off between resolution and computational cost.

\bibliographystyle{mn2e}
\bibliography{paper-mnras}

%

\label{lastpage}

\end{document}